\documentclass[sn-aps]{sn-jnl}%
\newcommand{\bv}[1]{\mathbf{#1}}		

\newcommand{\jbar}{\bar{j \phantom{\tiny \,}} \kern -0.1em}

\newcommand{\jhat}{\hat{j \phantom{\tiny \,}} \kern -0.1em}

\newcommand{\jtil}{\tilde{j \phantom{\tiny \,}} \kern -0.1em}

\newcommand{\vj}{\bv{j}}

\newcommand{\vv}{\bv{v}}

\newcommand{\vjbar}{\bar{\vj \phantom{\tiny \,}} \kern -0.1em}

\newcommand{\vjhat}{\hat{\vj \phantom{\tiny \,}} \kern -0.1em}

\newcommand{\vjtil}{\tilde{\vj \phantom{\tiny \,}} \kern -0.1em}

\newcommand{\Acal}{\mathcal{A}}

\usepackage{bm}

\newcommand{\vAcalhat}{\hat{\bm{\Acal} \phantom{a}} \kern-0.5em}
\newcommand{\vAcalcheck}{\check{\bm{\Acal} \phantom{a}} \kern-0.5em}
\newcommand{\vAcalbar}{\bar{\bm{\Acal} \phantom{a}} \kern-0.5em}

\usepackage{upgreek}

\usepackage{ulem}
\usepackage{graphicx}%
\usepackage{multirow}%
\usepackage{amsmath,amssymb,amsfonts}%
\usepackage{amsthm}%
\usepackage{mathrsfs}%
\usepackage[title]{appendix}%
\usepackage{xcolor}%
\usepackage{textcomp}%
\usepackage{manyfoot}%
\usepackage{booktabs}%
\usepackage{algorithm}%
\usepackage{algorithmicx}%
\usepackage{algpseudocode}%
\usepackage{listings}
\usepackage{comment}
\usepackage{subfloat}
\usepackage{soul}
\usepackage{textgreek}

\newcommand{\subscr}[2]{{#1}_{\textup{#2}}}

\theoremstyle{thmstyleone}%

\theoremstyle{thmstyletwo}%

\theoremstyle{thmstylethree}%
\newtheorem{definition}{Definition}%

\unnumbered

\begin{document}

\title[Article Title]{Charge-Density-Wave  Oscillator Networks for Solving Combinatorial Optimization Problems}

\author[1,2]{\fnm{Jonas Olivier} \sur{Brown}}\email{jonasbrown@ucla.edu}
\equalcont{These authors contributed equally to this work.}

\author[3]{\fnm{Taosha} \sur{Guo}}\email{tguo023@ucr.edu}
\equalcont{These authors contributed equally to this work.}

\author*[3]{\fnm{Fabio} \sur{Pasqualetti}}\email{fabiopas@ucr.edu}

\author*[1,2,4]{\fnm{and Alexander A.} \sur{Balandin}}\email{balandin@seas.ucla.edu}

\affil[1]{\orgdiv{Department of Materials Science and Engineering}, \orgname{University of California}, \city{Los Angeles}, \postcode{90095}, \state{California}, \country{USA}}

\affil[2]{\orgdiv{California NanoSystems Institute}, \orgname{University of California}, \city{Los Angeles}, \postcode{90095}, \state{California}, \country{USA}}

\affil[3]{\orgdiv{Department of Mechanical Engineering}, \orgname{University of California}, \city{Riverside}, \postcode{92507}, \state{California}, \country{USA}}

\affil[4]{\orgdiv{Center for Quantum Science and Engineering}, \orgname{University of California}, \city{Los Angeles}, \postcode{90095}, \state{California}, \country{USA}}


\abstract{

Many combinatorial optimization problems fall into the non-polynomial time NP-hard complexity class, characterized by computational demands that increase exponentially with the size of the problem in the worst case. Solving large-scale combinatorial optimization problems efficiently requires novel hardware solutions beyond the conventional von Neumann architecture. We propose an approach for solving a type of NP-hard problem based on coupled oscillator networks implemented with charge-density-wave condensate devices. Our prototype hardware, based on the 1\textit{T} polymorph of TaS\textsubscript{2}, reveals the switching between the charge-density-wave electron-phonon condensate phases, enabling room-temperature operation of the network. The oscillator operation relies on hysteresis in current-voltage characteristics and bistability triggered by applied electrical bias. This work presents a network of injection-locked, coupled oscillators whose phase dynamics follow the Kuramoto model and demonstrates that such coupled quantum oscillators naturally evolve to a ground state capable of solving combinatorial optimization problems. The coupled oscillators based on charge-density-wave condensate phases can efficiently solve NP-hard Max-Cut benchmark problems, offering advantages over other leading oscillator-based approaches. The nature of the transitions between the charge-density-wave phases, distinctively different from resistive switching, creates the potential for low-power operation and compatibility with conventional Si technology.}


\maketitle
The limitations of Moore’s law and rising energy demands in AI training have driven the development of unconventional computing methods to efficiently solve combinatorial optimization problems—common in real-world applications like routing, scheduling, and telecommunications—which can be mapped into Ising models \cite{ising1924beitrag, lenz1920beitrag}. Such problems belong to the NP-hard or NP-complete complexity classes. Finding the ground states of Ising models using digital computers requires exponentially growing resources as the problem size increases, making it difficult to solve even problems of moderate size efficiently.
Solving the Ising model efficiently requires novel hardware computation paradigms beyond conventional von Neumann architecture \cite{schuman2022opportunities, cai2020power}. Ising machines offer a promising approach to addressing complex and time-consuming optimization problems using traditional digital computing methods and algorithms \cite{mohseni2022, wang2021solving, Dutta2021, moy20221}.
Ising machines have been proposed on various technological platforms, including quantum-based systems like adiabatic quantum computing and quantum annealing using superconducting qubits \cite{Johnson2011, barends2016, puri2017quantum, Farhi2001, dickson2013, zhao2024}, CMOS-based systems such as digital and mixed-signal CMOS annealers \cite{Tsukamoto2017, yin2024ferroelectric, Yamaoka2016, singh2024cmos}, and optical systems utilizing coherent networks of degenerate optical parametric oscillators \cite{Marandi2014, roques2020, babaeian2019}. While these approaches show potential, they face significant challenges. Quantum annealers incur high operational costs and require complex cryogenic environments. Digital CMOS annealers struggle to maintain true randomness and require substantial post-processing. Optical coherent Ising machines (CIM) need long fiber ring cavities for temporal multiplexing and rely on power-intensive field-programmable gate arrays FPGAs for coupling. 
Annealing, quantum solutions, and dynamical system solvers are three approaches to addressing complex optimization problems. Simulated annealing (SA) has broad applicability, extending beyond Ising models to a wide range of optimization problems. However, it can be trapped in local optima, especially in complex energy landscapes, and requires many iterations to find the global minimum \cite{kirkpatrick1983optimization}.
Quantum solutions and dynamical system solvers use parallelism to explore multiple solutions simultaneously. Quantum entanglement enables quantum systems to navigate the energy landscape more efficiently. However, current quantum devices have limited qubit counts and are prone to errors, scalability issues, problems with generalizing to other optimization problems, and require complex cryogenic environments, leading to high operational costs \cite{Johnson2011, Farhi2001}. Dynamical system solvers deliver fast convergence, scale efficiently, and run on various hardware platforms. However, effectiveness can vary depending on the specific structure of the problem. In contrast, to slow annealing methods that maintain thermal equilibrium or ground state conditions, faster dynamical system approaches such as those using coherent Ising machines (CIMs) \cite{Marandi2014} or coupled oscillators \cite{wu1995application, Dutta2021, Lee2022, Zahedinejad2020, Mallick2020, Bashar2020, Ahmed2021, wu1998graph} drive the system toward the lowest energy state of the Ising model \cite{ising1924beitrag, lenz1920beitrag}. Optical CIMs require long fiber optic cables to process multiple signals over time and rely on power-intensive field-programmable gate arrays (FPGAs) for coupling, limiting scaling, and power efficiency \cite{Marandi2014}.
Recent studies \cite{wang2019oim, wang2015design, wang2021solving} have demonstrated the feasibility of using coupled oscillator networks to solve large-scale Ising problems. Experimentally, coupled electronic oscillator networks, such as coupled LC circuits, ring oscillators \cite{wu1998graph, wu1995application}, insulator-to-metal transition (IMT) oscillators \cite{Dutta2021, Lee2022}, spin oscillators \cite{Zahedinejad2020}, and integrated complementary metal oxide semiconductor (CMOS) oscillators, have been shown to solve Ising problems. However, there is a lack of theoretical explanation and guidance on designing coupled oscillator networks to generate the desired phase dynamics that efficiently solve Ising problems. If the phase dynamics of the oscillator network follow the Kuramoto model \cite{kuramoto1975self}, under proper injection locking, its phase dynamics evolve to a ground state that minimizes the Ising cost function 
\cite{wang2015design}. The stability properties of the Kuramoto model’s equilibrium points and the impact of injection signal strength on optimization performance are also analyzed in \cite{bashar2023stability, cheng2024control}.

In this article, we propose and demonstrate an operational coupled oscillator network based on charge-density-wave (CDW) devices, {whose phase dynamics follow the Kuramoto model.}
 The CDW phases are macroscopic quantum states characterized by a periodic modulation of the electronic charge density, accompanied by a small periodic distortion of the atomic lattice  \cite{GrunerDensitySolids, monceau2012electronic, zaitsev2004finite, balandin2021charge}. The CDW ground state is a condensate of electrons that differ in momentum by 2k\textsubscript{F}, or, in the equivalent interpretation, it is a condensate of 2k\textsubscript{F} phonons (k\textsubscript{F} is the Fermi wave vector). The early models for CDW phases and CDW transport included explicitly quantum mechanical  \cite{Bardeen1979} or classical descriptions \cite{Fisher1983, Fisher2005}. Certain phenomena observed in CDW materials, such as quantum creep \cite{ZaitsevZotov1993, CohnZaitsevZotov2023}, quantum interference \cite{zaitsev2004finite}, CDW Aharonov-Bohm effect \cite{Latyshev1997}, quantum tunneling \cite{Matsukawa1999}, and CDW quantization \cite{Zybtsev2010}, require a specific quantum mechanical description. The CDW phases have also been described as ordered quantum fluids that form in layered 1D or 2D van der Waals materials \cite{Miller2012, Miller2013}. While some aspects of CDW phenomena can be understood within classical or semi-classical models, one can consider the CDW a fundamentally quantum object. The CDW wavelength is nearly half the de Broglie wavelength of the Fermi-energy electrons, $\lambda = \frac{\pi}{k_f}$. The CDW distortion with this wavelength groups the electrons in pairs, resulting in the energy gap of $2\Delta$ in the electron spectrum at the Fermi level. Owing to the new CDW periodicity superimposed on the crystal lattice, the long phase coherence of the CDW, and the band-gap opening that affects electron transport, a 2D thin film in the CDW phase can be considered a type of quantum well superlattice. In this structure, the modulation of the electron wave function results in the formation of energy gaps.

To illustrate the CDW condensate approach in computing, we experimentally tested the building blocks of the oscillator network – coupled CDW devices with 1\textit{T}-TaS\textsubscript{2} channels – at room temperature. In the next step, we designed the Ising solver based on a resistively coupled CDW oscillator network, using the relative phase dynamics of each device. Applying second-harmonic injection locking locks the phase of each device into a pair of binary values representing Ising spins. The resistive coupling between each device can represent the connectivity matrix in the Ising cost function. We demonstrate a CDW oscillator network designed so its phase dynamics follow a behavior similar to the Kuramoto model. 
To demonstrate the feasibility and efficiency of the proposed CDW oscillator-based Ising solver, we simulate a six-by-six oscillator network to solve Max-Cut benchmark problems. The simulation results show that our oscillators converge to optimal solutions within 10 \textmu s. Thus, using strongly correlated CDW phase transitions instead of conventional resistive switching devices opens the possibility of low-power, scalable information processing for specific types of NP-hard problems.

\subsection{Two-Dimensional CDW Condensate Materials}

Several quasi-one-dimensional (1D) and quasi-two-dimensional (2D) metals reveal the strongly correlated CDW condensate phase \cite{GrunerDensitySolids, monceau2012electronic, zaitsev2004finite, balandin2021charge}. These phases unlock new device functionalities via voltage-triggered current non-linearities, the appearance of AC components under DC bias, hysteresis, and other phenomena \cite{balandin2021charge, Brown2023CurrentFilms, Taheri2022ElectricalDevices, taheri2023electric}. Interest in CDW condensate phases rapidly intensified after the realization that some phases persist at relatively high temperatures, enabling room-temperature (RT) operation of CDW devices. 
The 1\textit{T} polymorph of TaS\textsubscript{2} (1\textit{T}-TaS\textsubscript{2}) is a prominent member of the quasi-2D CDW materials, which reveal three CDW condensate phases the commensurate (C-CDW), nearly commensurate (NC-CDW) and incommensurate CDW phases (IC-CDW)
\cite{Thompson1971, manzke1989phase, Brown2023CurrentFilms}. The CDW phase transitions emerge when changing temperature (see Fig \ref{Fig: Figure_1}\textcolor{blue}{a}) or electrical bias and are accompanied by hysteresis in the current-voltage (I-V) characteristics (see Fig \ref{Fig: Figure_1}\textcolor{blue}{b}) \cite{Taheri2022ElectricalDevices, Brown2023CurrentFilms}. The hysteretic NC-CDW – IC-CDW phase transition, which occurs above RT (see Fig. \ref{Fig: Figure_1}\textcolor{blue}{a}) and is close to the operation temperatures of modern electronics, makes CDW quantum materials promising candidates as voltage-controlled oscillators (VCOs) \cite{Liu2016ATemperature}.

CDW-based quantum oscillators (CDW-QOs) derive functionality from phase transitions where the material's electronic properties undergo significant changes due to the rearrangement of charge density within the crystal lattice \cite{butler2020mottness}. The associated displacement of the atoms is small. For this reason, CDW devices can function at low power and operate at high speeds \cite{Liu2016ATemperature, mohammadzadeh2021evidence}. 
These considerations explain our motivation to search for alternative strongly correlated quantum materials that can provide the functionality required for solving NP-hard problems while simultaneously ensuring potentially low-power dissipation and fast operation. More details comparing other approaches for solving NP-hard problems to CDW-QOs are provided in Section \ref{sec: additional benefits}.
\section{Experimental and Simulated Results}

\subsection{Electrical Characteristics}
\begin{figure}[h]\label{Fig: Figure1.png}
\centering 
\includegraphics[width=\textwidth]{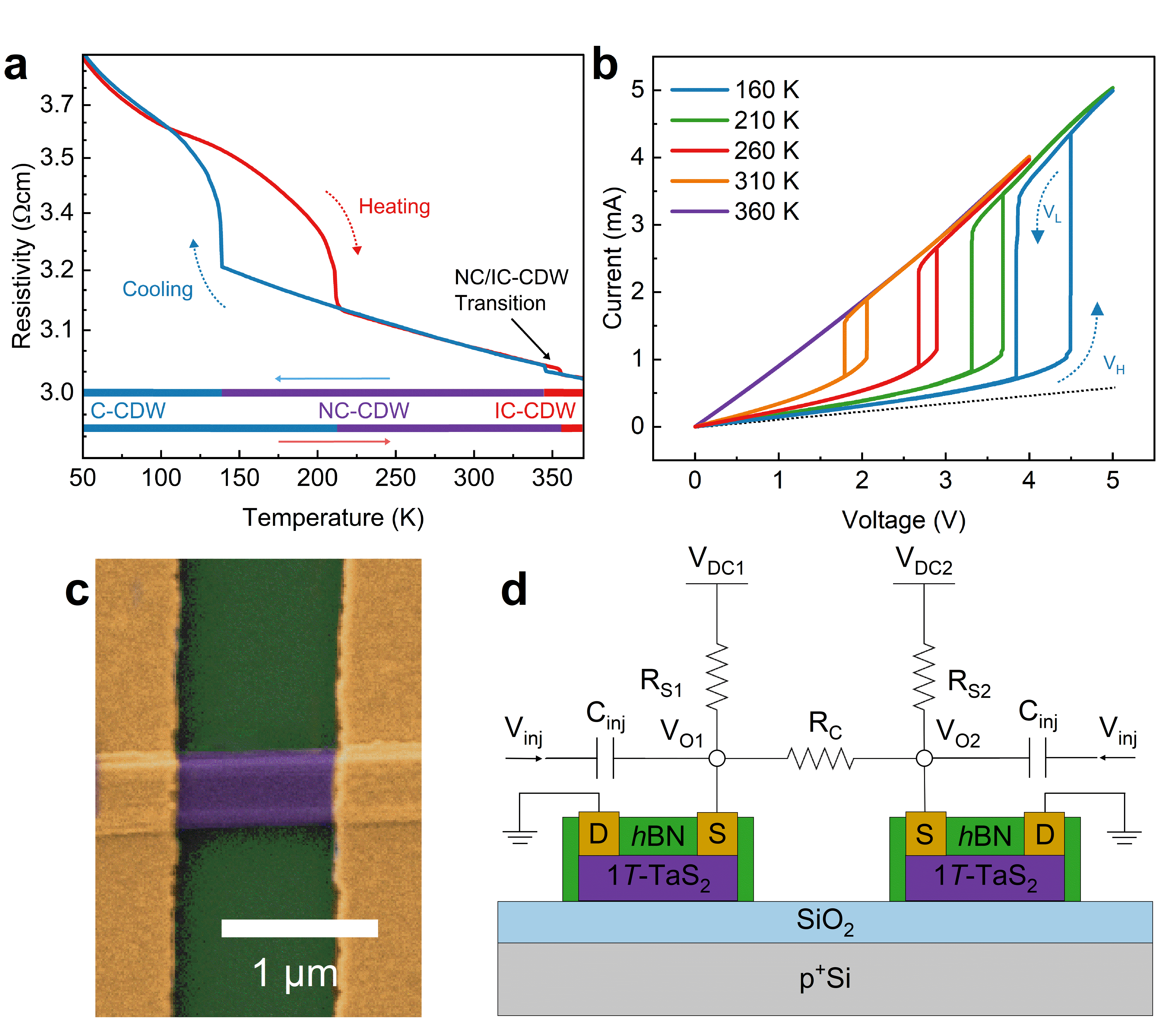}
\caption{\textbf{$|$ Electrical characteristics and circuit configuration of 1\textit{T}-TaS\textsubscript{2} CDW devices.} \textbf{a}, Resistivity versus temperature for a CDW device with a $\sim 100$-nm-thick channel. \textbf{b}, Temperature-dependent I-V characteristics of a 1\textit{T}-TaS\textsubscript{2} CDW device. \textbf{c}, SEM image showing a $\sim 1 $ \textmu m-long 1\textit{T}-TaS\textsubscript{2} device channel in the coupled oscillator circuit. The pseudo colors are used for clarity. \textbf{d}, Circuit schematic of the coupled oscillator device with off-chip coupling resistor R\textsubscript{C}, and load resistors R\textsubscript{Si}. The two devices are powered by an applied DC bias voltage V\textsubscript{DCi}, and all circuit elements are connected to a common ground. The injection locking signal is applied to both devices through an injection capacitor C\textsubscript{inj}, and an oscilloscope monitors the output V\textsubscript{Oi}.} 
\label{Fig: Figure_1}
\end{figure}

The temperature-dependent I-V characteristics of a representative device cooling in steps of 50 K from 360 K to 160 K, with a channel thickness of approximately 15 nm and channel width $\sim 2$ \textmu m, are displayed in Fig. \ref{Fig: Figure_1}\textcolor{blue}{b}. 
The hysteresis associated with the NC-CDW – IC-CDW transition exhibits a shape and size consistent with previous reports \cite{Brown2023CurrentFilms, yu2015gate, sipos2008mott, geremew2019proton, stojchevska2014ultrafast}. 
The I-V curve becomes super-linear before the initial current jump in the hysteresis region due to Joule heating induced by local heating from the current passage through the channel \cite{Liu2016ATemperature}. The self-heating effect, dependent on the device parameters and voltage sweep rate, induces the NC-CDW – IC-CDW phase transition in these devices \cite{Brown2023CurrentFilms, yu2015gate, sipos2008mott, geremew2019proton, stojchevska2014ultrafast}. 
Furthermore, inducing CDW phase transitions via an electric field rather than local heating is feasible \cite{Taheri2022ElectricalDevices, yu2015gate, Ravnik2021AMaterial, Mraz2022ChargeSpeed, zhu2018light}. Fig. \ref{Fig: Figure_1}\textcolor{blue}{b} illustrates the NC-CDW to IC-CDW phase transition as $\subscr{V}{H}$ and the IC-CDW to NC-CDW transition on the reverse bias as $\subscr{V}{L}$. After several months of testing, the fabricated 1\textit{T}-TaS\textsubscript{2} devices exhibited robustness with reproducible I-V characteristics. 
Fig. \ref{Fig: Figure_1}\textcolor{blue}{a} shows the resistance as a function of the temperature measured with a low bias (0.25 mV) for a thicker device $\sim 100$  nm. 
These transitions are consistent with prior reports for devices with similar thickness and channel width \cite{Brown2023CurrentFilms, sipos2008mott, wang2020band, yoshida2014controlling, yoshida2015memristive}.

\subsection{Charge-Density-Wave Based Quantum Oscillators}
\begin{figure}[h]
\centering
\includegraphics[width = \textwidth]{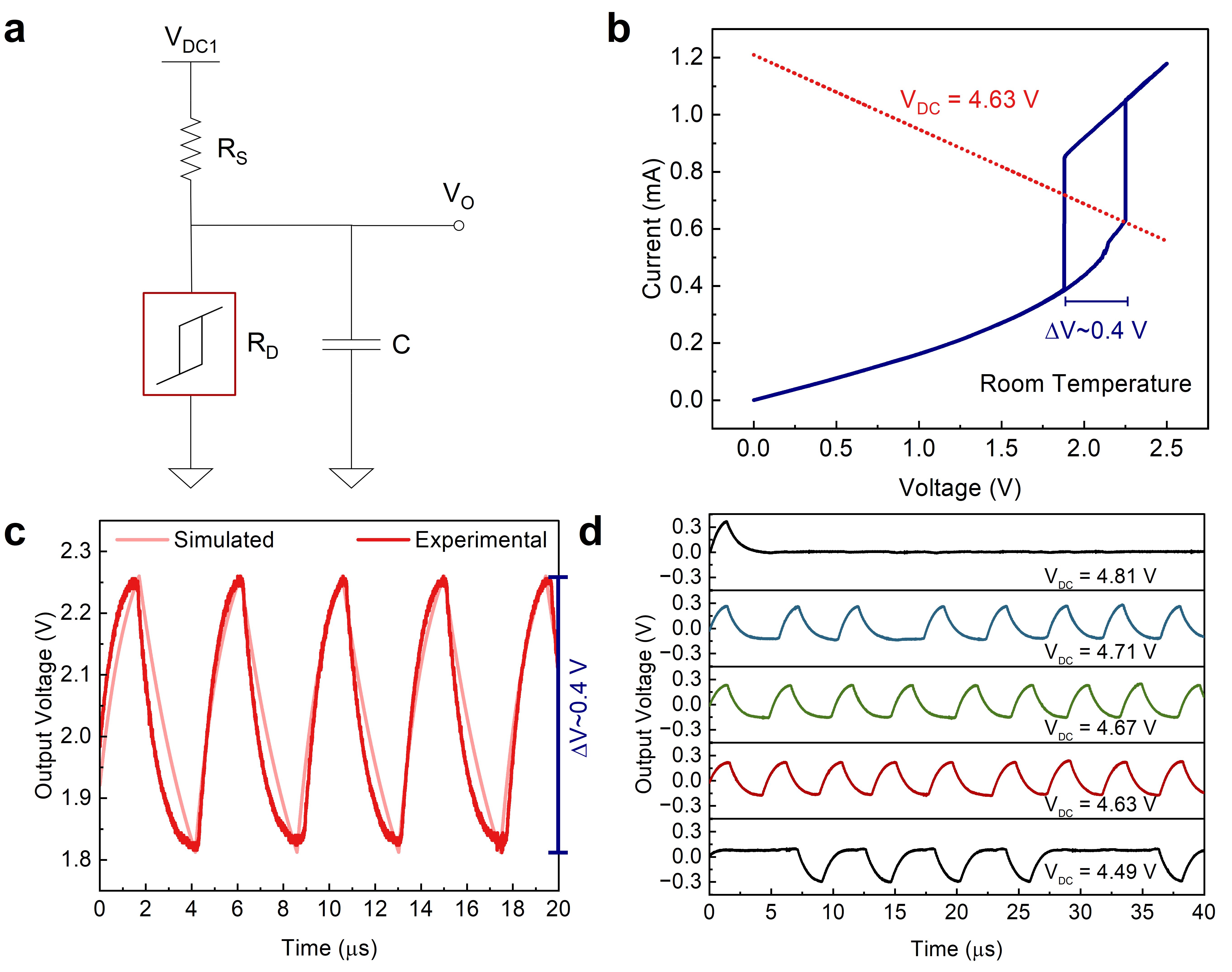}
\caption{\textbf{$|$ Oscillatory Characteristics of a CDW Device at room temperature. a}, Circuit schematic of a single CDW oscillator device consisting of 1\textit{T}-TaS\textsubscript{2} channel, an off-chip load resistor R\textsubscript{S}, and a lumped capacitance C from the output node V\textsubscript{O} to ground. An oscilloscope monitors the output terminal. \textbf{b}, The hysteresis in the I-V characteristics of a 1\textit{T}-TaS\textsubscript{2} device at room temperature. A resistive load line from the circuit described in \textbf{a} intersects the hysteresis. The width of the hysteresis is $\Delta V \sim 0.4$ V. \textbf{c}, Oscillations produced by the circuit shown in \textbf{a}. These oscillations only occur when the voltage across the device intersects the hysteresis, as shown in \textbf{b}. Pink traces represent simulated oscillations, and red traces indicate experimental results obtained at room temperature. \textbf{d}, The frequency of stable oscillations changes as the applied DC bias is adjusted. At $V_{DC} = 4.49$ V, the oscillations begin. Increasing to $V_{DC} = 4.63$ V, $4.67$ V, and $4.71$ V, oscillations stabilize with $f = 219$ kHz, $208$ kHz,
and $195$ kHz. At $V_{DC} = 4.81$ V, the oscillations become unstable. 
 The output voltage is normalized.} 
\label{Fig: Figure_2}
\end{figure}

The hysteresis exhibited during the phase transition in 1\textit{T}-TaS\textsubscript{2} enables the generation of voltage-controlled oscillations. This AC characteristic arises from the bistable resistance at a threshold bias voltage along the 1\textit{T}-TaS\textsubscript{2} channel. Fig. \ref{Fig: Figure_2}\textcolor{blue}{a} shows the circuit used to make a single CDW-QO device. 
Fig. \ref{Fig: Figure_2}\textcolor{blue}{b} shows the intersection of the resistive load line with the I-V hysteresis produced by the device at RT when the DC bias is $4.63$ V with $R_S = 2.26$ kΩ. 
Fig. \ref{Fig: Figure_2}\textcolor{blue}{c} shows the oscillations produced by the CDW-QO. The experimental results match the simulated results from the analytical expressions discussed in equation \eqref{eq: v dynamics} and equation \eqref{eq: v dynamics2}. 
Varying $\subscr{V}{DC}$ while maintaining a constant load resistance ($\subscr{R}{S} = 2.26 $ $k\Omega$) enables tuning of the oscillation frequency by adjusting the load line–hysteresis intersection as shown in Fig. \ref{Fig: Figure_2}\textcolor{blue}{d} and described in \eqref{eq: charging and discharging time}.

\subsection{Voltage dynamics of the CDW quantum oscillators}
{We determine the voltage dynamics of a CDW quantum oscillator using device hysteresis, the resistive load line, and the value for parasitic capacitance.}
Here  $\subscr{R}{H}$  is the resistance value during the high resistance state, and $\subscr{R}{L}$ is the resistance value during the low resistance state. Let $\subscr{R}{S}$ denote the circuit's load resistance and $C$ the parasitic capacitance of the device. Then, the output voltage ${\subscr{V}{O}}(t)$ across the device follows:
\begin{align}{\label{eq: v dynamics}}
\begin{cases} 
{\subscr{\dot{V}}{O}}(t) =  - \frac{1}{\subscr{R}{ch} C } \subscr{V}{O}(t) + \frac{1}{\subscr{R}{ch} C} \subscr{V}{ch} \;\; (charging),\\
{\subscr{\dot{V}}{O}}(t) =  - \frac{1}{\subscr{R}{dis} C } \subscr{V}{O}(t) + \frac{1}{\subscr{R}{dis} C} \subscr{V}{dis}\;\; (discharging),
\end{cases}
\end{align}
 where $\subscr{R}{ch} = \frac{\subscr{R}{H} \subscr{R}{S}}{\subscr{R}{H}+\subscr{R}{S}}$ and  $\subscr{R}{dis} = \frac{\subscr{R}{L} \subscr{R}{S}}{\subscr{R}{S}+\subscr{R}{L}}$ are the equivalent resistances for charging and discharging states, respectively;   $\subscr{V}{ch} = \frac{\subscr{R}{H}}{\subscr{R}{H}+\subscr{R}{S}}  \subscr{V}{DC} $ and 
$\subscr{V}{dis} = \frac{\subscr{R}{L}}{\subscr{R}{L}+\subscr{R}{S}}  \subscr{V}{DC} $ are the corresponding equivalent voltages for the charging and discharging states. 
We denote $\subscr{V}{H}$ and $\subscr{V}{L}$ as the highest and lowest voltage across a device, respectively.  Since the oscillator switches from charging to discharging when $\subscr{V}{O}(t) \geq \subscr{V}{H}$, and it switches from discharging to charging when $\subscr{V}{O}(t) \leq \subscr{V}{L}$, the solution  of equation \eqref{eq: v dynamics} is obtained as
\begin{align}{\label{eq: v dynamics2}}
\begin{cases}
\subscr{V}{O}(t) = (\subscr{V}{L} - \subscr{V}{ch}) e^{-t/(\subscr{R}{ch} C) } + \subscr{V}{ch}  \;\;  (charging), \\
\subscr{V}{O}(t) = (\subscr{V}{H} - \subscr{V}{dis}) e^{-t/(\subscr{R}{dis} C) } + \subscr{V}{dis} \;\; (discharging), \\
\end{cases}
\end{align}
and the charging time $\subscr{T}{ch}$ and discharging time $\subscr{T}{dis}$ within one oscillation cycle are:
\begin{align}{\label{eq: charging and discharging time}}
\begin{cases}
\subscr{T}{ch}  = - \frac{1}{\subscr{R}{ch} C} \log{\frac{\subscr{V}{H}-\subscr{V}{ch}}{\subscr{V}{L} -\subscr{V}{ch}}}, \: \\ \:
\subscr{T}{dis}  = - \frac{1}{\subscr{R}{dis} C} \log{\frac{\subscr{V}{L}-\subscr{V}{dis}}{\subscr{V}{H} -\subscr{V}{dis}}}. 
\end{cases}
\end{align}
For the sake of simplicity, in the following context, we use $f(\subscr{V}{O}(t))$ to represent the voltage dynamics of a free-running oscillator as in equation \eqref{eq: v dynamics}.

Connecting the outputs of individual oscillators with a resistive component creates a network of oscillators
(see Fig. \ref{Fig: Figure_3}\textcolor{blue}{a}). Pairs of resistively coupled CDW oscillator devices are investigated experimentally with numerical studies and analytical simulation. Coupling two devices with the correct coupling resistor, $\subscr{R}{C}$, allows the oscillator frequency to lock when the natural frequency of the oscillators is comparable. Fig. \ref{Fig: Figure_3}\textcolor{blue}{a} shows the circuit used to couple two CDW oscillator devices. The input voltages $\subscr{V}{DC1}$ and $\subscr{V}{DC2}$, as well as the load resistances $\subscr{R}{S1}$ and $\subscr{R}{S2}$ are selected to ensure stable oscillation states for the individual CDW-QOs. We tune the load resistors ${\subscr{R}{S}}_i$ individually for each 1\textit{T}-TaS\textsubscript{2} device, using the average resistance between the IC-CDW and NC-CDW resistive states. Adjusting ${\subscr{V}{DC}}_i$, ${\subscr{R}{S}}_i$, and $\subscr{R}{C}$ allows for various phase relationships between the oscillators. Before coupling, CDW-QOs oscillate at their natural frequencies, as shown in the lower panel of Fig. \ref{Fig: Figure_3}\textcolor{blue}{d}. The intersection of the load line with the hysteresis of the device determines the oscillation dynamics. The coupling resistance $\subscr{R}{C}$ is determined based on the desired coupling strength.
A higher coupling resistance yields weaker coupling; a lower resistance yields stronger coupling. Strong coupling aligns the oscillators in phase, while weak coupling shifts them out of phase. Fig. \ref{Fig: Figure_3}\textcolor{blue}{d} demonstrates this concept. The bottom panel shows oscillators before coupling where the oscillators are oscillating at their natural frequencies $f_1 = 537 $ kHz (red) and $f_2 = 476 $ kHz (blue). 

We assume that the differences between the oscillators can be neglected, namely $f(\subscr{V}{O$i$}(t))  \approx f(\subscr{V}{O$j$}(t)), \; \forall \; i, j \in \{1,2,\cdots,N\}$. When the $i$th oscillator is coupled with other oscillators through the coupling resistance $R_{ij}$, with $ j = \{1, 2, \cdots, N \}$,  its voltage follows
\begin{align} {\label{eq: coupled voltage dynamics}}
\subscr{\dot{V}}{O$i$}(t) =  f(\subscr{V}{O$i$}(t)) \underbrace{- \sum_{j=1}^{N}  \frac{1}{R_{ij}C} (\subscr{V}{O$i$}(t) - \subscr{V}{O$j$}(t))}_{\text{coupling term}},
\end{align}
where coupling is driven by the currents $I_{ij}(t) = \frac{1}{R_{ij}}(\subscr{V}{O$i$}(t) - \subscr{V}{O$j$}(t))$ flowing  from the $i$-th oscillator to other oscillators connected through the coupling resistance $R_{ij}$. 
{
We assume the oscillator network is weakly coupled, meaning that the $R_{ij}$ are chosen to be sufficiently large such that the currents  $I_{ij}(t)$ are small.}
{Let $\theta_i (t)$ denote the phase of the $i$-th oscillator. When the oscillator is free running, we have $\dot{\theta}_i (t) =  \omega_0$, where $\omega_0 = 2\pi/(\subscr{T}{ch}+\subscr{T}{dis})$ is the natural angular frequency. Using phase reduction theory \cite{nakao2016phase}, we show that the  phase dynamics of the coupled oscillator network in  equation \eqref{eq: coupled voltage dynamics} can be approximated by
\begin{align}
    \dot{\theta}_i(t) \approx \omega_0  -\sum_{j=1}^{N} J_{ij} \sin(\theta_i (t) - \theta_j (t) ), 
\end{align}
where  $J_{ij} = A/(R_{ij}C)$ and $A$ is a constant. We provide the derivation of the phase dynamics of the coupled CDW oscillator network in ``Methods".
We neglect differences between oscillators, assign each oscillator the same natural frequency $\omega_0$, and use $\phi_i(t) = \theta_i(t)- \omega_0t$ as the relative phase for each oscillator.
}

\section{Solving combinatorial optimization problems using CDW-QO networks}
Physicists first studied the Ising model in the 1920s as a mathematical framework to explain domain formation in ferromagnets \cite{ising1924beitrag, lenz1920beitrag}. The model comprises $N$ variables $\bm{x}= \begin{bmatrix}
    x_1 & x_2 & \cdots & x_N
\end{bmatrix}$, where each variable $x_i$ represents a spin that takes values in $\{+1,-1\}$ and minimizes an ``energy function," known as the Ising Hamiltonian  function:
\begin{align}\label{eq: Ising Hamiltonian}
    H(\bm{x}) =  -\sum_{1<i<j<N} W_{ij} x_ix_j, \; \text{with} \;  x_i, x_j \in \{-1,+1\},
\end{align} 
where  $W_{ij}$ are real coefficients.  The Ising model’s appeal stems from its ability to map real-world optimization problems directly to determine the ground-state solution of equation \eqref{eq: Ising Hamiltonian} \cite{lucas2014ising}. However, such problems belong to the NP-hard or NP-complete complexity classes.
Computing their solutions with digital computers requires exponentially growing resources as the problem size increases, making it difficult to solve problems of moderate size efficiently.

{To represent the Ising spins $x_i, x_j \in \{-1,+1\}$ defined in the Ising Hamiltonian cost function using $\{\phi_i(t)\}$, we {need to} lock $\{\phi_i(t)\}$  to a pair of binary values. Previous work \cite{todri2021frequency,neogy2012analysis,bhansali2009gen} demonstrates that by injecting an external periodic signal with an appropriate magnitude to each device, $\{\phi_i(t)\}$  can be locked to a certain phase configuration. To do so, the frequency of the injected signal should be an integer multiple of the device's natural frequency. In our settings, we choose a sinusoidal injection signal, $\subscr{V}{inj}(t) = \subscr{A}{inj}\sin(n\omega_0 t)$, where $\subscr{A}{inj}$ is the magnitude of the injected signal and $n$ is a positive integer. }

\begin{figure}[h]
\centering
\includegraphics[width=\textwidth]{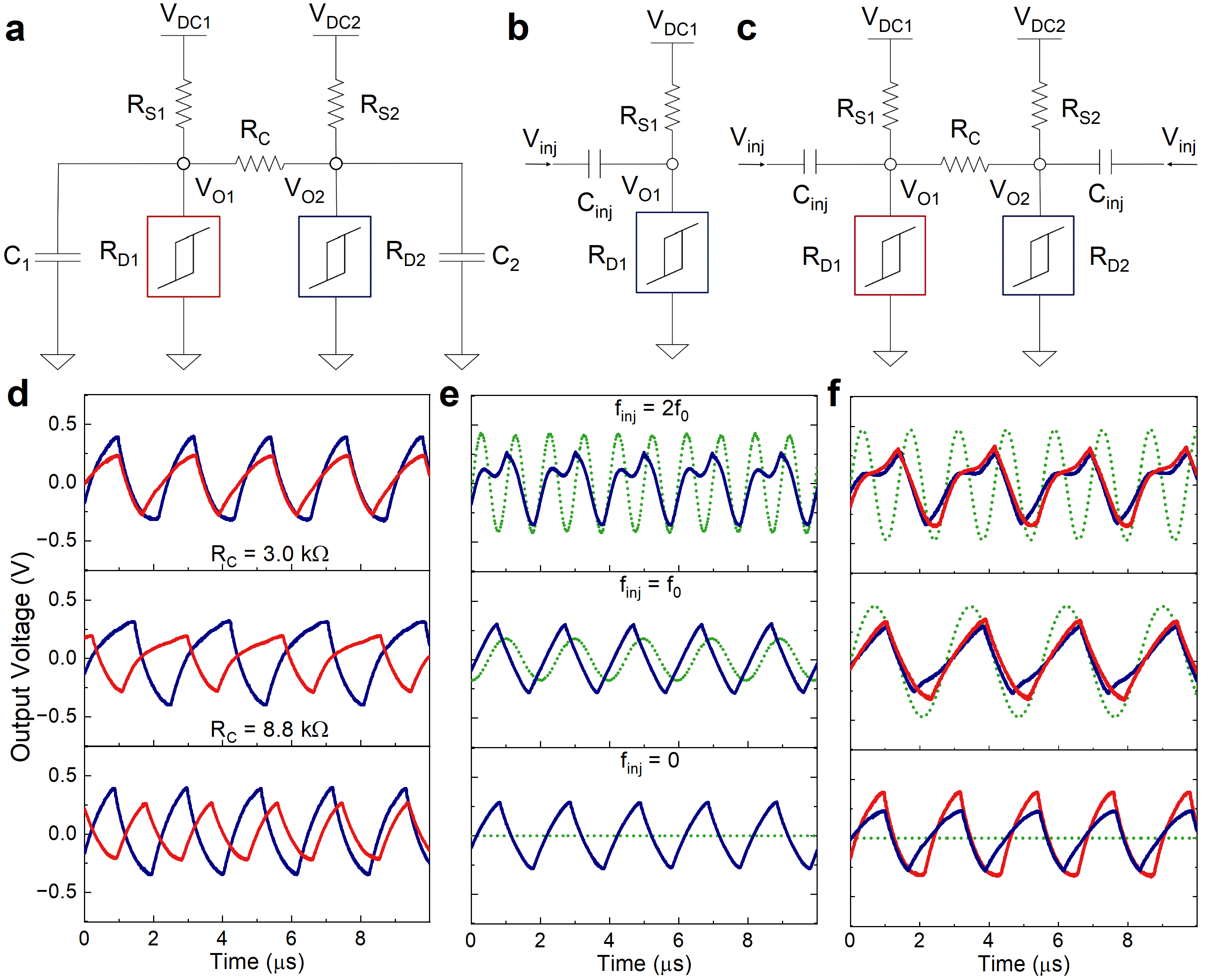}
\caption{\textbf{$|$ Oscillatory dynamics of coupled and injection-locked CDW-QOs.} \textbf{a}, Circuit describing two resistively coupled CDW oscillator devices. \textbf{b}, Circuit describing the injection locking of a single CDW-QO. \textbf{c}, The injection locking signal is applied to two resistively coupled CDW-QOs. \textbf{d}, Coupling scenarios for a pair of coupled CDW-QOs described by the circuit in \textbf{a}. Adjusting the coupling resistance tunes the phase of the frequency-locked coupled oscillators. A larger coupling resistance results in a smaller coupling between oscillators. \textbf{e} Injection locking scenarios for a single oscillator described by the circuit in \textbf{b}. The green dotted line represents the injection signal. Three scenarios are depicted (bottom to top): before injection locking, FHIL, and SHIL after annealing to the stable solution. \textbf{f} Injection locking scenarios of coupled CDW-QOs as described by the circuit in \textbf{c}. Three scenarios are depicted (bottom to top): Strong coupling before injection locking, after FHIL, and after SHIL. Note the output voltage is normalized for \textbf{d,e, and f}.} 
\label{Fig: Figure_3}
\end{figure}

{The schematic circuit of a single CDW-QO is shown in Fig. \ref{Fig: Figure_3}\textcolor{blue}{b}, where an injection signal $\subscr{V}{inj}$ is applied to the output $\subscr{V}{O1}$ through the injection capacitor $\subscr{C}{inj}$.  The circuit describing the pair of coupled CDW oscillators with injection locking is shown in Fig. \ref{Fig: Figure_3}\textcolor{blue}{c}. }
As shown in \cite{todri2021frequency,neogy2012analysis,bhansali2009gen}, the case of first-harmonic injection locking (FHIL) emerges when the frequency of $\subscr{V}{inj}(t)$ equals $\omega_0$, i.e., $\subscr{V}{inj}(t) = \subscr{A}{inj}\sin(\omega_0 t)$, $\{\phi_i(t)\}$ can be locked to the phase of the injection signal in the steady state.  When the frequency of $\subscr{V}{inj}(t)$ is twice the natural frequency $\omega_0$, i.e., $\subscr{V}{inj}(t) = \subscr{A}{inj}\sin(2\omega_0 t)$,  $\{\phi_i(t)\}$ can be locked to a pair of binary values in the steady state. {This corresponds to the case of second harmonic injection locking (SHIL).} {Fig. \ref{Fig: Figure_3}\textcolor{blue}{e} shows the experimental response of a CDW-QO to an injection signal. Here, the bottom panel shows the free-running oscillator, the middle panel shows the response of the CDW-QO to FHIL, and the top panel shows the response of the oscillator to SHIL. The free-running oscillator has an infinite number of stable phase configurations; under FHIL, the CDW-QO has only one stable phase configuration, and with SHIL, there are two stable phase configurations. It should be noted that, for the case of the free running and the SHIL, only one stable phase configuration is shown in Fig. \ref{Fig: Figure_3}\textcolor{blue}{e}.}


{For the case of coupled CDW-QOs, Fig. \ref{Fig: Figure_3}\textcolor{blue}{c} shows the circuit describing injection locking the network.} Fig. \ref{Fig: Figure_3}\textcolor{blue}{f} {demonstrates the three injection locking scenarios of a pair of coupled CDW-QOs experimentally. The bottom panel shows a pair of strongly coupled oscillators (R\textsubscript{C} = 3 k$\Omega$) before injection locking. The middle panel shows the coupled oscillators under FHIL, where the phase of both oscillators synchronize with the injection signal. The top panel shows one of the bistable configurations of the CDW-QOs under SHIL.}

Three different injection-locking scenarios are experimentally demonstrated in the context of two weakly coupled oscillators in Fig. \ref{Fig: Figure_4}. Similar to the central panel of Fig. \ref{Fig: Figure_3}\textcolor{blue}{d}, {Fig. \ref{Fig: Figure_4}\textcolor{blue}{a} shows the oscillations formed from weak coupling without injection locking.  Fig. \ref{Fig: Figure_4}\textcolor{blue}{b} shows the FHIL of a coupled pair of CDW-QOs, where we linearly anneal the injection amplitude to $\subscr{A}{inj} = 2 $ V. For the case of SHIL, each oscillator exhibits bistability, meaning they will settle into one of the two phase-locked states, effectively binarizing the phase. Fig. \ref{Fig: Figure_4}\textcolor{blue}{c} shows one of the two stable phases of the oscillators where the oscillators are in phase with one another. {  Fig. \ref{Fig: Figure_4}\textcolor{blue}{d} shows the other stable phase where the CDW-QOs are out of phase with one another.} The experimental results for these three scenarios have been confirmed with numerical simulations, as shown in light blue and light red in Fig. \ref{Fig: Figure_4}. 


\begin{figure}[h]
\centering
\includegraphics[width=0.75\textwidth]{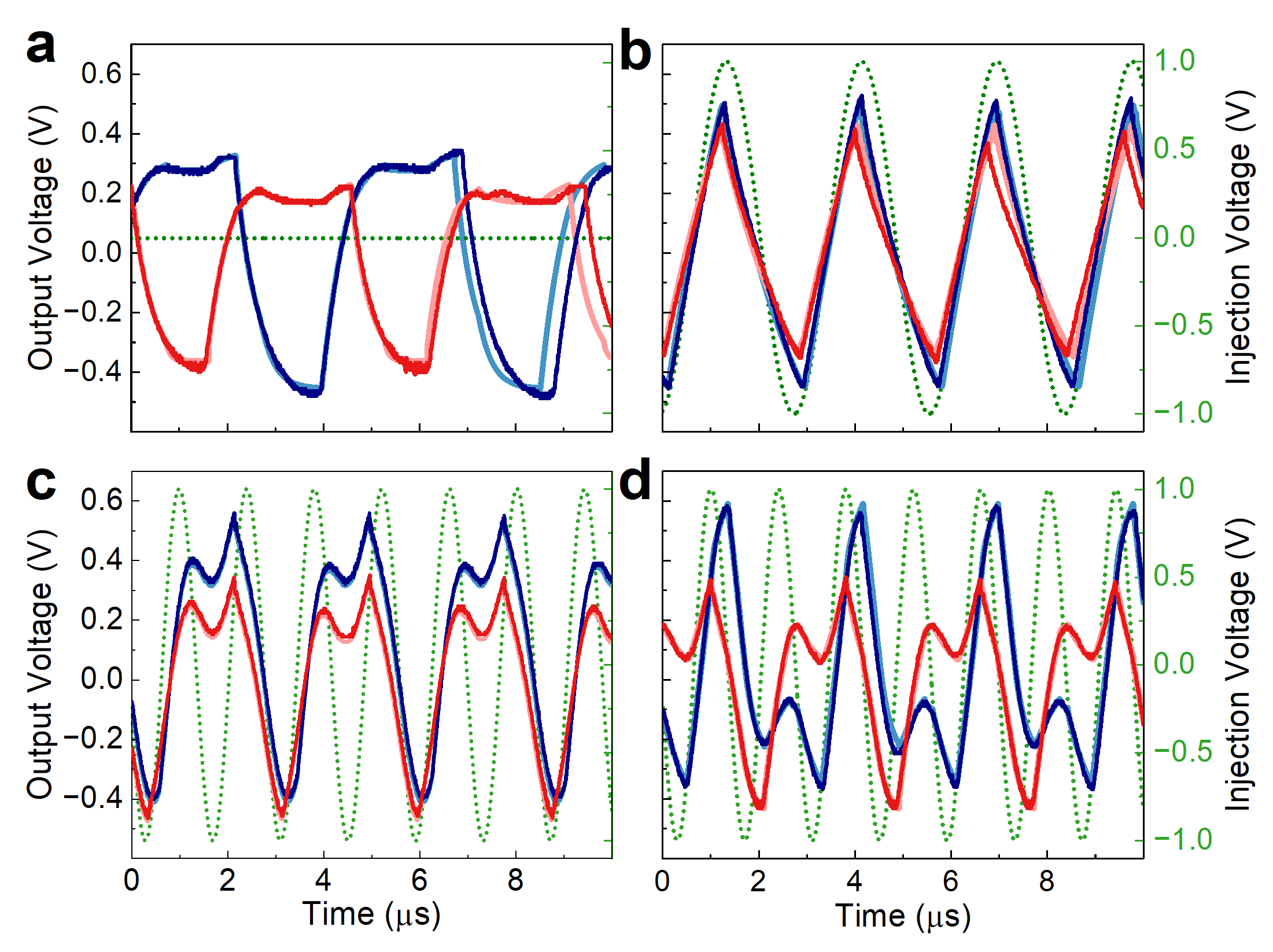}
\caption{\textbf{Experimental injection-locking scenarios}. \textbf{a}, In the absence of injection locking, weakly coupled oscillators synchronize in frequency but not in phase, resulting in a flat energy landscape for the phase space. \textbf{b}, With FHIL, the oscillators synchronize their phases with the injection locking signal. \textbf{c,d}, For SHIL, the oscillator phase is in a bi-stable state and has in-phase \textbf{c}, and out-of-phase \textbf{d}, configurations to the input with equal probability. The green dotted line shows the injection locking signal. Simulated results, shown in light blue and light red, demonstrate the oscillators fit the theoretical framework.} 
\label{Fig: Figure_4}
\end{figure}

{
We also simulate the scenarios of FHIL and SHIL for a pair of homogeneous coupled CDW-QOs.  For FHIL, the voltage waveforms of the two oscillators are in-phase and locked to the input signal (see Fig. \ref{Fig: Figure_5}\textcolor{blue}{b}). 
Then we simulate their phase dynamics $\phi_1(t)$ and $\phi_2(t)$ using equation \eqref{eq: phase dynamics with inj} for
$10$ experiments with randomly generated initial values between $0^{\circ}$ and $360^{\circ}$. We notice that  $\phi_1(t)$ and $\phi_2(t)$, starting from different initial values, are all locked to $117^{\circ}$ in the steady state, as shown in Fig. \ref{Fig: Figure_5}\textcolor{blue}{e}.
For SHIL,  the voltage waveforms of the
two oscillators exhibit in-phase
and out-of-phase configuration with respect to the input signal, as shown in Fig. \ref{Fig: Figure_5}\textcolor{blue}{c}.
We also simulated their phase dynamics $\phi_1(t)$ and $\phi_2(t)$ following equation \eqref{eq: phase dynamics with inj} 
for $20$ experiments with randomly generated initial values.  We observe that $\phi_1(t)$ and $\phi_2(t)$ starting from different initial values converge to a pair of binarized phase values: $112^{\circ}$ and $292^{\circ}$ (see Fig. \ref{Fig: Figure_5}\textcolor{blue}{f}). For the case of zero injection signal, we simulate the phase evolution of $\phi_1(t)$ and $\phi_2(t)$ for 10 experiments with randomly generated initial values.  As shown in Fig. \ref{Fig: Figure_5}\textcolor{blue}{f}, they settle to many different values spread out in the phase domain.}

{Under SHIL, the oscillator phase responses  $\{\phi_i(t)\}$ are locked to a pair of binary phase values $\{112^{\circ}, 292^{\circ}\}$ differing by $180^{\circ}$, which we use to represent the Ising spins $\{-1,+1\}$  in equation \eqref{eq: Ising Hamiltonian}. This phenomenon is demonstrated in various types of oscillators  \cite{neogy2012analysis, wang2015design}. }
The bi-stability of $\phi(t)$ offers a method for encoding the Ising spin within the electrical domain, and mapping the bi-stable behavior to the Ising spins allows for solving combinatorial optimization problems such as the Max-Cut. 
The simulated results demonstrate the oscillators fit the theoretical framework and thus can be used to scale up the problem.
{A comprehensive analysis of the coupled CDW oscillator network’s phase dynamics $\{\phi_i\}$ is provided in “Method,” which also discusses how SHIL can be used to lock $\{\phi_i\}$ to a pair of values that can represent Ising spins.}

\begin{figure}[h]
\centering
\includegraphics[width=\textwidth]{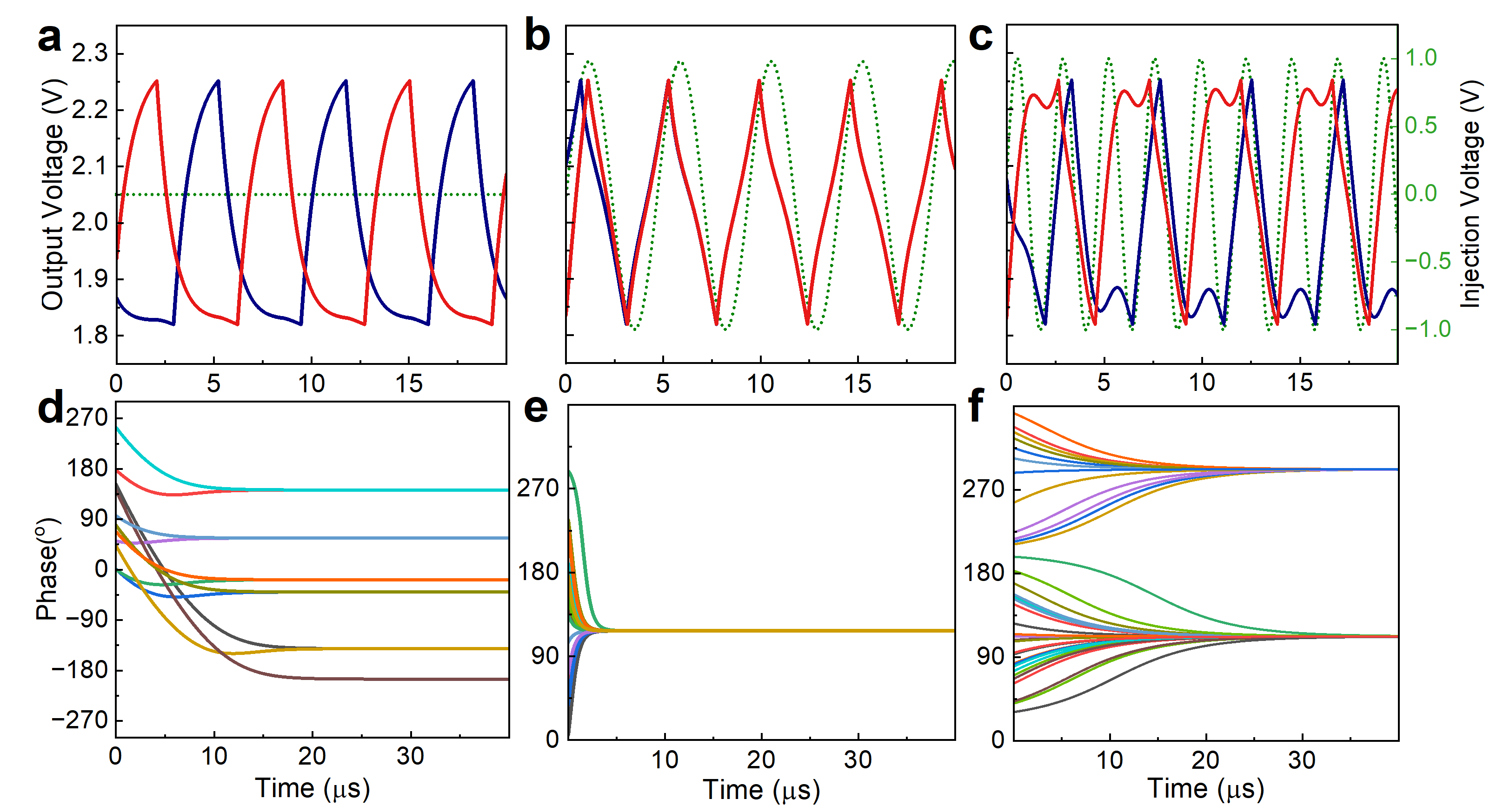}

\caption{\textbf{Phase evolution of three distinct injection-locking scenarios}. \textbf{a,d}, In the absence of synchronization, the free-running oscillators exhibit uniform phases across the phase space. \textbf{b,e}, For FHIL, the oscillators are in-phase and locked with the injection signal. Their corresponding phases $\phi_1(t)$ and  $\phi_2(t)$ with different random initial values converge to the locked phase value $117^{\circ}$. \textbf{c,f}, For SHIL, the oscillators exhibit in-phase and out-of-phase steady-state waveforms locked to the injection signal, and $\phi_1(t)$  and $\phi_2(t)$ with random initial values converge to the bi-stable phase values $112^{\circ}$  and  $292^{\circ}$ with  $50\%$ probability respectively.}  
\label{Fig: Figure_5}
\end{figure}

\subsection{Benchmark examples in solving the Max-Cut problem}
We demonstrate our CDW oscillator network's feasibility and efficacy by solving a class of Ising Hamiltonian problems: The maximum cut problem (Max-Cut). Max-Cut is one of the most straightforward graph partitioning problems to conceptualize, yet one of the most challenging combinatorial optimization problems to solve \cite{floudas2008encyclopedia}. Its objective is to partition the set of vertices of a graph into two subsets, such that the sum of the weights of the edges having one endpoint in each of the subsets is maximized. The proper definitions of a graph and a cut are given as follows:
\begin{definition}
Let $\mathcal{G} = (V, E)$ be an un-directed graph, where V is the vertex set and E is the edge set. Let $W := \{W_{ij}\}$ be the weights
of each edge in the graph. A cut is defined as a partition of the vertex set into two disjoint subsets $S$ and $V \backslash S$. The weight of the cut $(S, V \backslash S)$ is given by 
\begin{align}
   C(S): = \sum_{i\in S, j \in V\backslash S } W_{ij}.
\end{align}
\end{definition}
\begin{definition}
    The maximum cut of the graph $\mathcal{G} = (V,E)$ with weight $W$ is defined as  
    \begin{align}{\label{eq: maximum cut}}
    C^{*}(S): = \max_{\forall S\subseteq V}   C(S) .
\end{align}
\end{definition}
The MAX$-$CUT problem seeks to find a subset of vertices such that the total weights of the cut set between this subset and the remaining vertices are maximized. Let $x_i = 1 $  \text{if} $ i \in S$ or $x_i = -1$ \text{if}$ \: i \in V \backslash S$,  then $C(S)$ can be written as
\begin{align}
  C(S) = \frac{1}{2} \sum_{1\leq i \leq j \leq n} W_{ij}(1- x_i x_j) =  \frac{1}{2} \sum_{1\leq i \leq j \leq n} W_{ij} - W_{i,j}x_ix_j.
\end{align}
Let $\bm{x} = \begin{bmatrix}
    x_1 & x_2 & \cdots & x_N
\end{bmatrix}$ and define $H(\bm{x}) = \sum_{1\leq i \leq j \leq n} W_{ij}x_i x_j $. If we choose $J_{ij} = -W_{ij}$, $H(\bm{x})$ is in the form of the Hamiltonian cost function in equation \eqref{eq: Ising Hamiltonian}. 
Therefore, maximizing the cut values $C(S)$ is equivalent to minimizing the  Hamiltonian cost function $H(\bm{x})$ with respect to $J_{ij} = -W_{ij}$. 

To investigate the performance of the CDW coupled oscillator network as Ising Hamiltonian solvers, in our simulation setup, we choose an undirected  and unweighted graph with  $6$ nodes  as shown in Fig. \ref{Fig: Figure_6}\textcolor{blue}{a}, with  connectivity (adjacency) matrix
{{
\begin{align}\label{eq: connectivity matrix}
     W = \frac{1}{C\subscr{R}{C}}
    \begin{pmatrix}
     0  &   1  &   1  &   1  &   0  &   0 \\
     1   &  0   &  1  &   1  &   1   &  1 \\
     1   &  1  &   0   &  1  &   0   &  1 \\
     1  &   1   &  1   &  0  &   1   &  1\\
     0   &  1  &   0   &  1  &   0   &  0\\
     0   &  1   &  1  &   1  &   0   &  0
    \end{pmatrix}.
\end{align}}} There are two sets of optimal  partitions to obtain the maximum cut value $C^{*}(S) = \frac{8}{C\subscr{R}{C}}$: 
\begin{align}\label{eq: optimal partition}
    x_2 = x_3 = x_4 =1, \; x_1 = x_5 = x_6 =-1, 
      \end{align}
      and 
    \begin{align}
    x_2 = x_4 =1, \; x_1 = x_3 = x_5 = x_6 =-1.
    \end{align}
Fig \ref{Fig: Figure_6}\textcolor{blue}{b} depicts the circuit illustration of our six-by-six coupled CDW oscillator network. Each oscillator connected to others through coupling resistances of equal magnitude  $\subscr{R}{C} = 43.727 $ $k\Omega$, following the connectivity matrix $W$ in equation \eqref{eq: connectivity matrix}. The same SHIL signal is applied to each oscillator through the injection capacitance $\subscr{C}{inj}$.

\begin{figure}[h] 
\centering
\includegraphics[width=\textwidth]{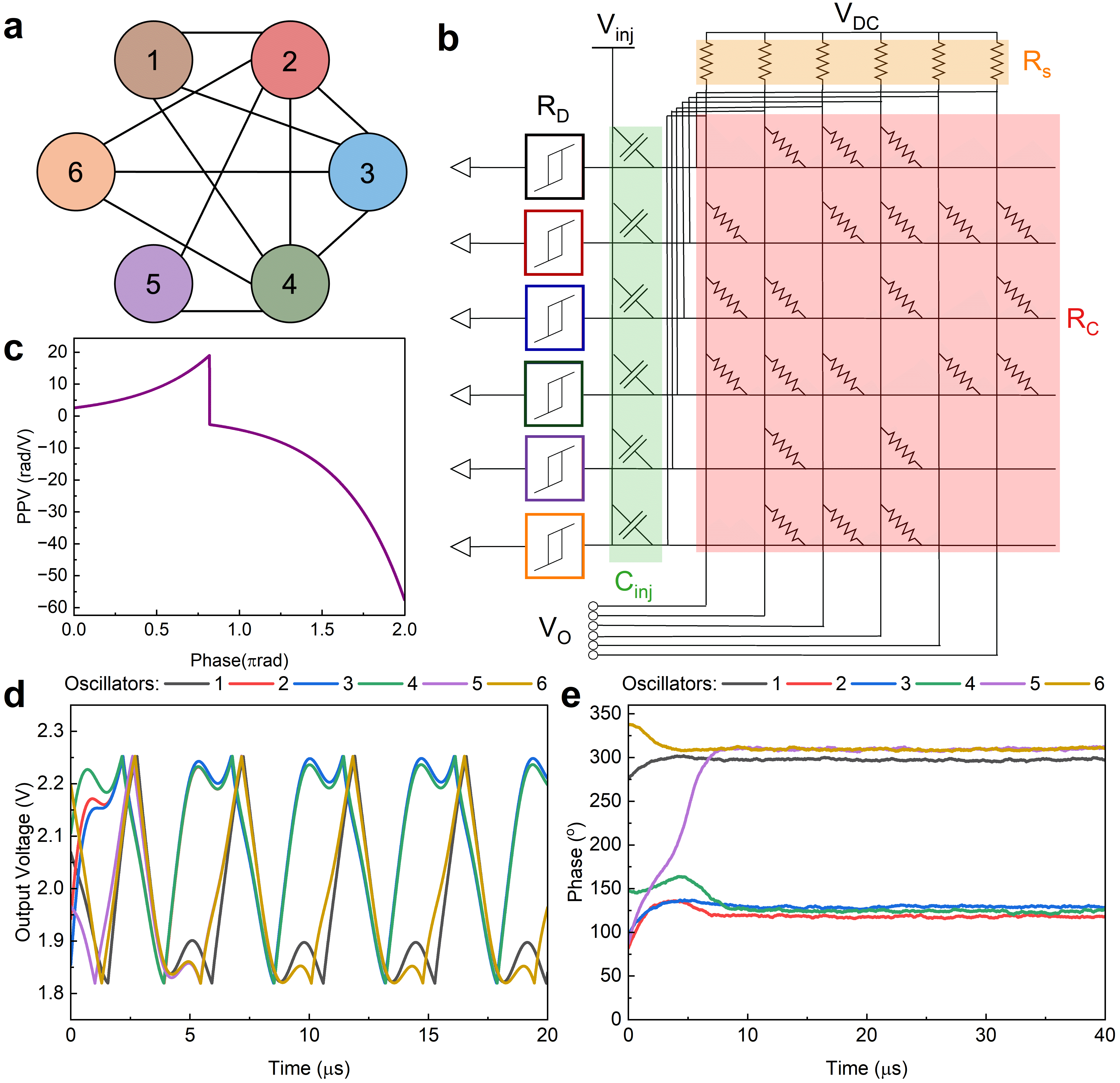}
\caption{The simulated results from solving Max-Cut optimization problems. 
\textbf{a}, The $6 \times 6$ connected graph, where the weight of all the edges equals $1/(C\subscr{R}{C})$. \textbf{b}, Circuit representation of the $6$ coupled oscillators using the weights described in the connectivity matrix. \textbf{c}, The values of the phase sensitivity function over $0$ to $2\pi$. \textbf{d}, The voltage waveforms of oscillators $1$, $5$, and $6$ are synchronized, while oscillators $2$ and $3$  are synchronized with oscillator $4$.
\textbf{e}, The phase evolution of the six-dimensional oscillator network. In the steady state $\phi_1$, $\phi_5$ and $\phi_6$ converge to  $115^{\circ}$, whereas $\phi_2$, $\phi_3$, and $\phi_4$  converge to $298^{\circ}$.} 
\label{Fig: Figure_6}
\end{figure}




Fig. \ref{Fig: Figure_6}\textcolor{blue}{d} and Fig. \ref{Fig: Figure_6}\textcolor{blue}{e} show our simulation results. We first randomly initialize the voltage values $\vv_i(t)$ for each oscillator.  As shown in Fig. \ref{Fig: Figure_6}\textcolor{blue}{c}, we observe that $\vv_2(t)$, $\vv_3(t)$ and, $\vv_4(t)$ are anti-phase synchronized  with $\vv_1(t)$, $\vv_5(t)$ and $\vv_6(t)$ after $t \approx 10$ \textmu s.  We also observe that their corresponding phases $\phi_2(t)$, $\phi_3(t)$ and $\phi_4(t)$ converge to $298^{\circ}$,
which is $180^{\circ}$ different from the phase value $115^{\circ}$ that $\phi_1(t)$, $\phi_5(t)$ and $\phi_6(t)$ converge to, as shown in Fig. \ref{Fig: Figure_6}\textcolor{blue}{e}. 
 Thus, the simulation results demonstrate that the steady states of the six coupled oscillators give one optimal partition that solves the Max-Cut problem for the graph in equation \eqref{eq: optimal partition}. 

\section{Methods}

\subsection{Fabrication}
Bulk 1\textit{T}-TaS\textsubscript{2} crystals were grown using the chemical vapor transport method (CVT)\cite{Binnewies2013ChemicalReview}. Thin layers of 1\textit{T}-TaS\textsubscript{2} were prepared by mechanical exfoliation and placed onto a SiO\textsubscript{2}/p\textsuperscript{+}Si substrate using an in-house transfer system. The ability to fabricate CDW-QOs on SiO\textsubscript{2} substrates highlights their seamless compatibility and potential for integration with existing Si CMOS technology. A thin \textit{h}-BN capping layer was applied atop the 1\textit{T}-TaS\textsubscript{2} layers to protect samples during nanofabrication and prevent environmental damage. Electron-beam lithography (EBL) was employed to fabricate two-terminal device structures. Atomic layer etching of the \textit{h}-BN with SF\textsubscript{6} plasma exposes the 1\textit{T}-TaS\textsubscript{2} flake, enabling deposition of Ti/Au (10 nm/100 nm) metal contacts via Electron Beam Evaporation (EBE). Fig. \ref{Fig: Figure_1}\textcolor{blue}{c} shows a false color SEM image of a two-terminal device with a channel width of $\sim 1 $ \textmu m. Fig. \ref{Fig: Figure_1}\textcolor{blue}{d} shows a schematic of a pair of devices used in this study. The schematic illustrates a pair of coupled 1\textit{T}-TaS\textsubscript{2} devices on a Si/SiO\textsubscript{2} substrate, covered with \textit{h}-BN, accompanied by off-chip load resistors $\subscr{R}{si}$, coupling resistance $\subscr{R}{C}$, parasitic capacitance $C_i$, injection signal $\subscr{V}{inj}$ , and injection capacitance $\subscr{C}{inj}$. Both devices are energized by a DC bias voltage $\subscr{V}{DC}$, with all circuit components connected to a shared ground. An oscilloscope reads the output voltage ${\subscr{V}{O$i$}}$, and we measure the phase of the oscillators under various coupling configurations.

\subsection{Electrical Measurements}
Temperature-dependent electrical characterization of the 1\textit{T}-TaS\textsubscript{2} devices was performed using a cryogenic probe station (Lakeshore TTPX) and a semiconductor parameter analyzer (Agilent B1500A). A cryogenic temperature controller (Lakeshore 336) maintained the environmental temperature. Heating and cooling rates were set at a constant 2 K/min using the Lakeshore Measure LINK software to ensure accurate temperature readings and preserve sample integrity. For the current-voltage (I-V) measurements, the applied bias voltage was swept across the device channel at a rate of 200 mV/s with a step size of 2 mV.
We use a Teledyne LeCroy WaveAce 1012 oscilloscope to capture the waveform's oscillating signal, while a BK Precision 1787B provides a constant DC voltage bias to the circuit.
The Teledyne Lecroy Wavestation 2012 waveform generator provides the injection locking signal to the oscillator circuit for both FHIL and SHIL scenarios.

\subsection{Phase dynamics of the coupled oscillators}
 When an oscillator connects with other oscillators in the network through the coupling resistance  $R_{ij}$, we have
\begin{align}
\dot \theta_i(t) = \omega_0 - \sum_{j=1}^{N}  \frac{1}{R_{ij}C} \Gamma (\theta_i (t) - \theta_j (t)),
\end{align}
where 
\begin{align}{\label{eq: Gamma function}}
\Gamma(\theta_i (t) - \theta_j (t)) = \frac{1}{2\pi}\int_{0}^{2\pi}Z(\theta_i(t) - \theta_j(t) + \psi) (\subscr{V}{O}(\theta_i(t) - \theta_j(t)+\psi) - \subscr{V}{O}(\psi))d \psi
\end{align} is the phase coupling function and $Z(\theta) = \frac{\partial \theta(t)}{\partial \subscr{V}{O}}$ is the $2\pi$ periodic phase sensitivity function \cite{nakao2016phase}. One straightforward  method to compute $Z(\theta)$ numerically is to use its definition: $Z(\theta) = \lim_{\Delta \subscr{V}{O}  \to 0} \frac{\Delta \theta }{\Delta \subscr{V}{O}}$. Specifically, we perturb the limit cycle voltage dynamics $f(\subscr{V}{O}(t))$  by adding $\Delta \subscr{V}{O} = 0.01$ over $1170$ steps covering one charging and discharging cycle. We compare the perturbed waveform $\subscr{\Tilde{V}}{O}(t)$ to the original waveform $\subscr{V}{O}$, and measure the time shift $\Delta T$ between their switching points. Then, the corresponding phase shift is calculated as $\Delta\theta = 2 \pi \Delta T/(\subscr{T}{ch}+\subscr{T}{dis})$.  Fig. \ref{Fig: Figure_6}\textcolor{blue}{c} shows the calculated phase sensitivity function waveform over the interval of $0$ and $2 \pi$. 

Following  equation \eqref{eq: Gamma function}, we numerically calculate the phase coupling function $\Gamma(\theta_i(t) - \theta_j(t))$ over $0$ to $2\pi$ and we further observe that $\Gamma(\theta_i(t) - \theta_j(t))$ can be approximated by a sinusoidal function as 
$A \sin (\theta_i(t)-\theta_j(t)+\alpha)+\beta$, with $A= -1.93$, $\alpha = 0.06439$ and $\beta= -0.1033$, see Fig. \textcolor{blue}{S4b}. This indicates that the phase response of the CDW-coupled oscillators follows dynamics similar to Kuramoto oscillators \cite{acebron2005kuramoto}. Since the values of $\alpha$ and $\beta $ are determined by the oscillator's voltage dynamics, it is possible to adjust the device temperature and circuit variables, such as the DC voltage and load resistance values,  to generate the desired phase coupling functions.
Other examples \cite{wang2014phlogon, lo2023ising} also show that many oscillators, such as LC and ring oscillators, have the flexibility to tune the dynamics to yield the desired phase coupling functions.

Assuming the differences between oscillators can be neglected and every oscillator has the same natural frequency $\omega_0$, we introduce the  relative phase: $\phi_i(t) = \theta_i(t)- \omega_0t$ for each oscillator, where  $\phi_i(t)$ obeys 
\begin{align}{\label{eq: relative phase}}
    \dot{\phi}_i(t) = \dot{\theta}_i(t) -\omega_0 = -
\sum_{j=1}^{N} \frac{1}{R_{ij}C}\Gamma (\phi_i (t) - \phi_j (t)).
\end{align}

Fig. \ref{Fig: Figure_3}\textcolor{blue}{c} shows the schematic circuit of coupled oscillators with the injection signal. Since $\subscr{V}{inj}(t)$ is applied through a capacitor $\subscr{C}{inj}$, the corresponding injected current is $\subscr{I}{inj}(t) = \subscr{C}{inj} \frac{d\subscr{V}{inj}(t)}{dt} = \subscr{C}{inj} n \omega_0 \subscr{A}{inj}  \cos(n\omega_0t)$.  Then, the coupled voltage dynamics in equation \eqref{eq: coupled voltage dynamics} with the injection signal can be modified as
\begin{align}
\dot{\vv}_i(t) =  f(\vv_i(t)) - \sum_{j=1}^{N}  \frac{1}{R_{ij}C}  (\vv_i(t) - \vv_j(t))+\frac{1}{C}\subscr{I}{inj}(t),
\end{align}
{and the phase dynamics of the CDW oscillator network with injection signals  becomes
\begin{align}\label{eq: phase dynamics with inj}
    \dot{\phi}_i(t) =  
    -\sum_{j=1}^{N} \frac{1}{R_{ij}C}\Gamma (\phi_i (t) - \phi_j (t) ) + \subscr{K}{s}{\underbrace{\int_{0}^{2\pi}Z(\phi_i(t)+\psi)cos(n\psi)d\psi}_{\text{injection term}},}
\end{align}}
where $\subscr{K}{s} = 1/2\pi\subscr{C}{inj}/C n \omega_0 \subscr{A}{inj} $.
To explore how to use the binary phase values to represent the spin assignments of the cost function in equation \eqref{eq: Ising Hamiltonian}, next, we analyze the dynamics of $\{\phi_i(t)\}$ under SHIL.
First, we numerically calculate the injection term in equation \eqref{eq: phase dynamics with inj} for $n=2$, i.e., the SHIL scenario for $\phi_i(t)$ over the interval of $0$ to $2\pi$ (see Supplemental Fig. \textcolor{blue}{S5c}, solid line). The injection term can be approximated using a sinusoidal function: $A_1\cos(2\phi_i(t)+\alpha_1)+\beta_1$,  with $A_1=4.05$, $\alpha_1 = -2.31$ and $\beta_1 = 0.0014$ (see Supplemental Fig. \textcolor{blue}{S5c}, dashed line). Since  the phase coupling function $\Gamma(\phi_i(t) - \phi_j(t))$ can be approximated by $A\sin(\phi_i(t) - \phi_j(t)+\alpha)+\beta$, where $\alpha$ and $\beta$ are  small, and the injection term in equation \eqref{eq: phase dynamics with inj} can be approximated by $A_1\cos(2\phi_i(t)+\alpha_1)+\beta_1$, where $\beta_1$ is also small,  the dynamics of $\phi_i(t)$ in  equation \eqref{eq: phase dynamics with inj} can be approximated as:
\begin{align}\label{eq: simplified phase dynamics}
    \dot{\phi}_i(t) \approx  -\sum_{j=1}^{N} J_{ij} \sin(\phi_i (t) - \phi_j (t) ) + K_s\cos(2\phi_i(t)+\alpha_1),
\end{align}
{
where $J_{ij} = A/(R_{ij}C)$ and $K= A_1\subscr{K}{s}$.
Let $\tilde{\alpha}_1 = \alpha_1/2 - \pi/4$, then $\dot{\phi}_{i}(t)$ can be rewritten as
\begin{align}
\dot{\phi}_i(t) = -\sum_{j=1}^N J_{ij} \sin(\phi_i(t) -\phi_j(t)) - K_s \sin(2\phi_i(t) + 2 \tilde{\alpha}_1),
\end{align} 
which is the gradient flow of the following function
\begin{align}{\label{eq: relaxed optimization}}
    -\sum_{1\leq i < j \leq N} J_{ij}\cos(\phi_i(t) - \phi_j(t)) + K_s \sin^2(\phi_i(t)+\tilde{\alpha}_1).
\end{align}
We observe that equation \eqref{eq: relaxed optimization} is the relaxed version, with penalty coefficient $K_s$, of the following optimization problem :
\begin{equation}{\label{eq: equivalent Ising}}
\begin{aligned}
   & \min_{\phi_1,\cdots,\phi_N} \;\;\;\;\; -\sum_{1\leq i < j \leq n} J_{ij}\cos(\phi_i(t) - \phi_j(t)) \\
 & \text{subject} \; \text{to} \;\;\;\;\;\; \sin(\phi_i(t)+\tilde{\alpha}_1)=0, \; \text{for} \; i\in\{1,\cdots,N\}. 
\end{aligned}
\end{equation}
The optimization problem given by equation \eqref{eq: equivalent Ising} is equivalent to the Ising problem in equation \eqref{eq: Ising Hamiltonian} because the constraint limits $\phi_{i}+ \tilde{\alpha}_1$ to be either $0$ or $\pi$, which leads to $\cos(\phi_i(t) - \phi_j(t)) \in \{1,-1\}$ for $i,j \in \{1, \cdots, N\}$.}

{
By carefully tuning the penalty coefficient $K_s$ of the constraint term in equation \eqref{eq: equivalent Ising}, we can lock $\phi_i(t)$ at the binary phase values $\{112^{\circ}, 292^{\circ}\}$ that correspond to  $\sin(\phi_i(t)+\tilde{\alpha}_1) = 0$ and $\cos(\phi_i(t) - \phi_j(t)) \in \{1,-1\}$, as observed in our simulation of two coupled oscillators under SHIL. 
However, some equilibrium points might exist ($\dot{\bm{\phi}}(t) = 0$) that correspond to local minima of equation \eqref{eq: equivalent Ising} that do not yield the optimal solution to the Ising Hamiltonian problem, as discussed in \cite{bashar2023stability,cheng2024control}. A comprehensive analysis of equilibrium points and their structural and stability properties is carried out in \cite{cheng2024control}. For more details, we refer readers to this source. }

Other experimental works \cite{todri2021frequency, avedillo2023operating} for VO\textsubscript{2} oscillators demonstrate that the local minima are numerically avoided by slowly increasing the injection strength (annealing). Applying annealing improves the success rates of  VO\textsubscript{2} oscillators in solving combinatorial optimization problems. One interesting future direction for our setups would be analyzing the energy landscape of the CDW oscillator's phase dynamics to provide theoretical design methods and guarantees to reduce the effects of local minima.

\section{Additional Benefits of CDW Oscillator Devices}{\label{sec: additional benefits}}

CDW oscillators with 1\textit{T}-TaS\textsubscript{2} channels can be implemented on SiO\textsubscript{2} / Si substrates, simplifying their integration with conventional Si CMOS technology. The demonstrated IMT devices, e.g. with VO\textsubscript{2} active layers, have utilized TiO\textsubscript{2} \cite{Dutta2021}, and conductive TiN \cite{Mian2015} substrates or were deposited on SiO\textsubscript{2} using pulsed laser deposition (PLD) and atomic layer deposition (ALD) \cite{Corti2020a}. The maximum frequency of oscillations reported for a VO\textsubscript{2} device on SiO\textsubscript{2} is 100 kHz \cite{Corti2020a}.  Furthermore, limitations in the frequency of oscillations achievable by VO\textsubscript{2}, as indicated by electrothermal simulations \cite{Carapezzi2023}, restrict its utility in applications requiring higher frequencies and hinder the necessary time to reach a stable solution. In this study, individual 1\textit{T}-TaS\textsubscript{2} devices have exhibited stable oscillations up to 3.3 MHz at room temperature with an amplitude of just 0.01 V, showcasing the versatility of 1\textit{T}-TaS\textsubscript{2} CDW-QOs across a broad frequency spectrum with low power output (see Fig. \textcolor{blue}{S4}).
Oscillatory neural networks (ONNs) encode information by maintaining a stable phase difference between each oscillator in the network and a reference oscillator, achieved during synchronization \cite{Corti2020, Hoppensteadt2000}. The computational energy consumed by ONNs is directly proportional to their settling time \cite{Delacour2023}, which is determined by the number of cycles required for synchronization multiplied by the oscillation period.
Increasing the oscillator frequency and amplitude of oscillations will reduce the overall energy consumption of CDW-QO operations.

Radiation-induced damage is a significant challenge in modern electronics \cite{Rathod2011RadiationReview}, particularly in the harsh environment of outer space \cite{Keys2008High-performanceEnvironments}. The performance of CMOS devices can significantly deteriorate under X-ray or proton irradiation \cite{Buehler1989RadiationMeasurements}. Total ionizing dose (TID) irradiation can create electron-hole pairs within the oxide layers and at the interfaces in field-effect transistor (FET) structures \cite{liu2017total, Frank2001}. The accumulation of charge carriers can significantly affect the electrical characteristics and overall performance of the CMOS device. This degradation underscores the need to develop radiation-hardened electronics that overcome CMOS technology’s physical limits by engineering circuits from radiation-resistant materials. VCOs operating at RT built using CDW phase transitions in quasi-2D 1\textit{T}-TaS\textsubscript{2} are a strong alternative \cite{Liu2016ATemperature}.
CDW materials, being metals with high electron concentrations, are inherently more radiation-resistant than semiconductors. The radiation hardness of 1\textit{T}-TaS\textsubscript{2} VCOs against X-rays has been experimentally confirmed \cite{liu2017total}.

Another aspect to consider is the difference between CDW phase transitions in 1\textit{T}-TaS\textsubscript{2} and IMT phase transitions in materials such as VO\textsubscript{2} \cite{PhysRevLett.3.34}. VO\textsubscript{2} undergoes a reversible polymorph phase transition from a semiconducting monoclinic phase to a metallic rutile phase at $\sim 340 $ K. 1\textit{T}-TaS\textsubscript{2} experiences a reversible CDW phase transition from the more resistive NC-CDW state to the less resistive IC-CDW state. In CDW phase transitions, the material's electronic properties undergo significant changes due to the rearrangement of charge density within the crystal lattice. The transition in 1\textit{T}-TaS\textsubscript{2} does not involve a polymorph change but rather a resistance transition due to the switching between CDW states.

\section{Conclusions}\label{sec13}

We demonstrated an approach for solving certain NP-hard problems using coupled oscillator networks implemented with CDW quantum condensate devices. We built and tested prototype hardware based on the 1\textit{T} polymorph of TaS\textsubscript{2}, which revealed the switching between the CDW electron-phonon condensate phases above room temperature. The oscillator operation relies on hysteresis in the I-V characteristics and bistability triggered by applied electrical bias. The designed injection-locked, coupled oscillator network reveals the phase dynamics that follow the Kuromoto model. We demonstrate that the phase dynamics of such coupled quantum oscillators evolve to the ground state, which solves combinatorial optimization problems. The coupled oscillators based on charge-density-wave condensate phases can efficiently solve NP-hard Max-Cut benchmark problems, offering advantages over other leading oscillator-based approaches. The quantum nature of the transitions between the CDW phases, distinctively different from resistive switching, creates the potential for low-power operation and compatibility with conventional Si technology.


\normalem
\bibliography{real}

\end{document}


\title[Article Title]{1T-TaS\textsubscript{2} Coupled Charge-Density-Wave Oscillators to Solve Optimization Problems
}

\title[Article Title]{Supplementary Information: Coupled Charge-Density-Wave Oscillators for Solving Combinatorial Optimization Problems}

\author[1]{\fnm{Jonas Olivier} \sur{Brown}}\email{jonasbrown@ucla.edu}
\equalcont{These authors contributed equally to this work.}

\author[2]{\fnm{Taosha} \sur{Guo}}\email{tguo023@ucr.edu}
\equalcont{These authors contributed equally to this work.}

\author[2]{\fnm{Fabio} \sur{Pasqualetti}}\email{fabiopas@engr.ucr.edu}

\author*[1,3]{\fnm{Alexander A} \sur{Balandin}}\email{balandin@seas.ucla.edu}

\affil[1]{\orgdiv{Materials Science and Engineering}, \orgname{University of California}, \city{Los Angeles}, \postcode{90024}, \state{CA}, \country{USA}}

\affil[2]{\orgdiv{Mechanical Engineering}, \orgname{University of California}, \city{Riverside}, \postcode{92507}, \state{CA}, \country{USA}}

\affil[3]{\orgdiv{California NanoSystems Institute}, \orgname{University of California}, \city{Los Angeles}, \postcode{90024}, \state{CA}, \country{USA}}
\maketitle

\section{Contents}
\subsection{Supplemental Note 1: Atomic Force Microscopy Image of a representative device}
\subsection{Supplemental Note 2: Temperature dependence of
the oscillation frequency}
\subsection{Supplemental Note 3: Tunable Phase of Coupled
CDW-QOs}
\subsection{Supplemental Note 4: First Harmonic Injection
Locking}
\subsection{Supplemental Note 5: Oscillatory Characteristics of a High-Frequency Low Power CDW-QO at RT}
\subsection{Supplemental Note 6: Proof of Theorem. 1}

\newpage
\subsection*{Supplemental Note 1: Atomic Force Microscopy Image of a Representative Device}
The Atomic Force Microscopy (AFM) image reveals the topography of a representative device. The 1\emph{T}-TaS\textsubscript{2} channel, has a thickness of \(\sim 40 \, \text{nm}\). A protective hexagonal boron nitride (hBN) capping layer, approximately \(30 \, \text{nm}\) thick, protects the channel from oxidation and other contaminants. The device features Ti / Au contacts with a combined thickness of \(\sim 110 \, \text{nm}\). Here \(\sim 10 \, \text{nm}\) of Ti is used as an adhesion layer to the SiO\textsubscript{2} and 1\emph{T}-TaS\textsubscript{2} channel and \(\sim 100 \, \text{nm}\) of Au is used to ensure efficient electrical conductivity. The channel width is \(\sim 1 \, \mu \text{m}\), which is significant for the device's scale and performance.
\begin{figure}[h]\label{Figures: SupFigure1.png}
\centering
\includegraphics[width=0.6\textwidth]{Figures/SupFigure1.png}
\caption*{\textbf{Fig. S1 $|$ AFM image showing the topography of a representative device.} The thickness of the 1\emph{T}-TaS\textsubscript{2} channel is $\sim 40$ nm, the thickness of the hBN capping layer is $\sim 30$ nm, and the thickness of the Ti / Au contacts is $\sim 110$ nm. The channel width is $\sim 1$ $\mu$m.} 
\end{figure}

\newpage
\section*{Supplemental Note 2: Temperature dependence of the oscillation frequency}
The intersection of the resistive load line with the hysteresis is directly related to the oscillation frequency and the charging and discharging times of the CDW-QOs. By maintaining a constant load resistance $R_{L} = 6.3$ $k\Omega$ and applied bias voltage $V_{DC} =  $ the resistive load line remains the same according to the circuit described in Fig. \textcolor{blue}{2a}. The temperature of the environment increases from $250$ K to $300$ K, thus reducing $V_{H}$ and the width of the hysteresis window similar to Fig. \textcolor{blue}{1b}. Fig. \textcolor{blue}{S2a} shows how the hysteresis changes with the change in temperature from 250 K to 300 K and how the resistive load line intersects the hysteresis. The shape and the frequency of oscillations depend on how the load line intersects the hysteresis.
The position of the intersection between the load line and the hysteresis curve influences the charging and discharging times of the system. When the intersection point is closer to the higher threshold voltage (V\textsubscript{H}) the discharging process occurs more rapidly. This is because the system has a greater voltage difference to overcome during the discharge phase, resulting in a faster transition from the high resistance to the low resistance state. In contrast, when the intersection point is closer to the lower threshold voltage (V\textsubscript{L}) the charging process is faster. In this scenario, the system requires less voltage change to reach the high state, which leads to a faster charging time. Fig. \textcolor{blue}{S2c} shows the parabolic dependence of the frequency of oscillations as the intersection of the load line with the hysteresis changes from $250$ K to $300$ K. These measurements show the device will produce stable oscillations over a large temperature range. Also, stable oscillations were observed over 3 hours after being turned on and the device continues to operate successfully over a year after its fabrication demonstrating the robust nature of the devices presented.

\begin{figure}[h]\label{Figures: SupFigure2.png}
\centering
\includegraphics[width=0.8\textwidth]{Figures/SupFigure2.png}
\caption*{\textbf{Fig. S2 $|$ Temperature dependence of the oscillation frequency.} \textbf{a}, I-V characteristics of a CDW-QO at various temperatures from 250 K to 300 K. The dotted black line is the load line similar to Fig. \textcolor{blue}{2b}. \textbf{b}, Maintaining the same resistive load line oscillations at various temperatures from 250 K to 300 K are recorded. Adjusting the temperature changes the frequency of the oscillations and shape. \textbf{c}, The frequency of oscillations as a function of the temperature. The error is derived from two sources: the Full Width at Half Maximum (FWHM) of the amplitude versus frequency plot of the Fast Fourier Transform (FFT) of the oscillations, and the uncertainty in temperature measurement.} 
\end{figure}

\newpage
\section*{Supplemental Note 3: Tunable Phase of Coupled CDW-QOs}
Adjusting the $V_{DCi}$, $R_{si}$ and $R_C$ will affect the oscillations produced. The oscillations of a pair of 1\emph{T}-TaS\textsubscript{2} oscillators with constant coupling resistance $R_{C} = 4$ $k\Omega$ are shown in Figure 4c. Here, the voltage applied to the individual devices is varied, demonstrating different phase configurations. As shown in the single oscillator case (Fig. \textcolor{blue}{3d}) varying the applied voltage tunes the natural frequency of the oscillators. When the oscillators are coupled, their frequencies become locked or synchronized. However, as the applied voltage is adjusted, the natural frequencies of the individual oscillators diverge from this synchronized state. This mismatch between their intrinsic frequencies and the locked frequency results in a phase difference emerging between the coupled oscillators. The degree of this phase difference is determined by the magnitude of the voltage adjustment and the resulting shift in the oscillators' natural frequencies relative to the locked frequency. The phase of the coupled oscillators can also be tuned using the coupling resistor. Figure 4d shows the effect of changing the coupling resistance from $1.94 k\Omega$ to $R_C = 4 k\Omega$ to $R_{C} = 4.77$ $k\Omega$ resulting in various phase configurations. Figure 4b shows how the phase of the oscillators changes when adjusting the coupling resistor $R_{C}$. When the coupling resistance is $R_C = 4$ $k\Omega$ the oscillators are $\sim 30^{\circ}$ out of phase. Decreasing the coupling resistance to $1.95$ $k\Omega$ the oscillators are $\sim 5.80^{\circ}$ out of phase and the coupling strength is greater. Increasing the coupling resistance to $4.77$ $k\Omega$ the coupling strength is less, and the oscillators are no longer in phase.  Figure 4b shows how the frequency of coupled oscillations is related to the phase of the oscillators to each other. The plot shows that the frequency is higher for oscillators in phase with one another and as the phase difference increases, the frequency of oscillations also decreases.

\begin{figure}[h]\label{Fig: Figure4.png}
\centering
\includegraphics[width=0.8\textwidth]{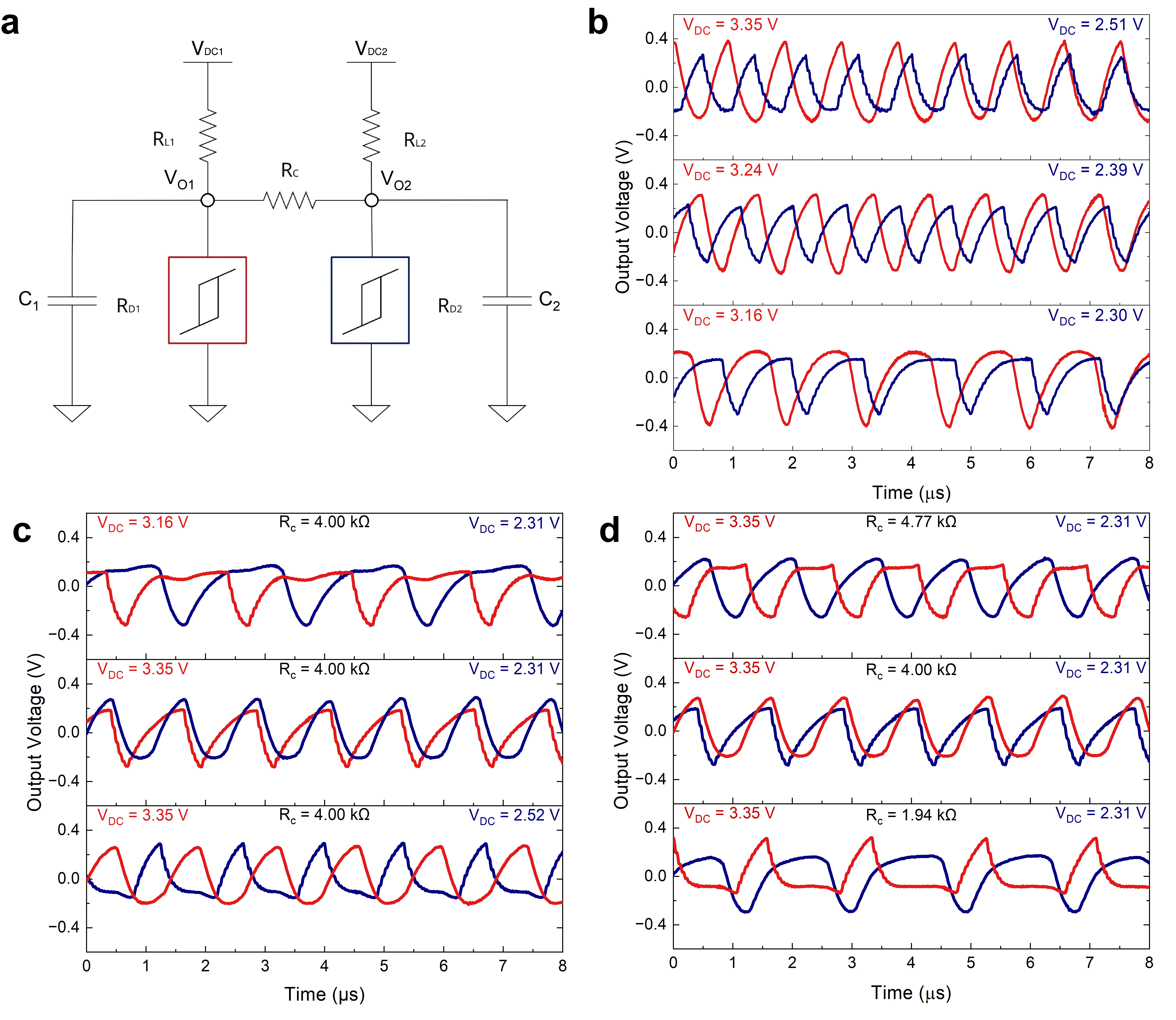}
\caption*{\textbf{Fig. S3 $|$ Tunable Phase of Coupled CDW-QOs.} \textbf{a}, Circuit describing two resistively coupled CDW oscillator devices. \textbf{b}, Oscillators before coupling have different natural frequencies. Adjusting the applied DC bias, V\textsubscript{DC} and load resistance R\textsubscript{S} tunes the frequency of the oscillators independently. \textbf{c}, Coupled oscillators with constant coupling resistance R\textsubscript{C} = 4.00 k$\Omega$. Varying the applied DC bias demonstrates the frequency locking of the coupled oscillators. The phase of the oscillators is tuned by adjusting V\textsubscript{DC} and R\textsubscript{S}. \textbf{d}, The phase of the frequency-locked coupled oscillators can also be tuned by adjusting the coupling resistance. A larger coupling resistance results in a smaller coupling between oscillators.} 
\end{figure}

\newpage
\section*{Supplemental Note 4: First Harmonic Injection Locking}
Fig. \textcolor{blue}{2c} shows how increasing the amplitude of the injected waveform (annealing) allows greater phase synchronization. Here the amplitude of the injection locking signal was linearly increased from $A_{inj} = 0$ mV to $400$ mV and snapshots at $A_{inj}= 5$ mV, $100$ mV,  and $200$ mV are shown. This helps to show how the CDW oscillator waveform evolves as the amplitude of the injection locking signal increases. The top two sections of Figure 5d show the effect of linearly increasing the amplitude of the injection signal to achieve better phase synchronization.
\begin{figure}[h]
\centering
\includegraphics[width=0.8\textwidth]{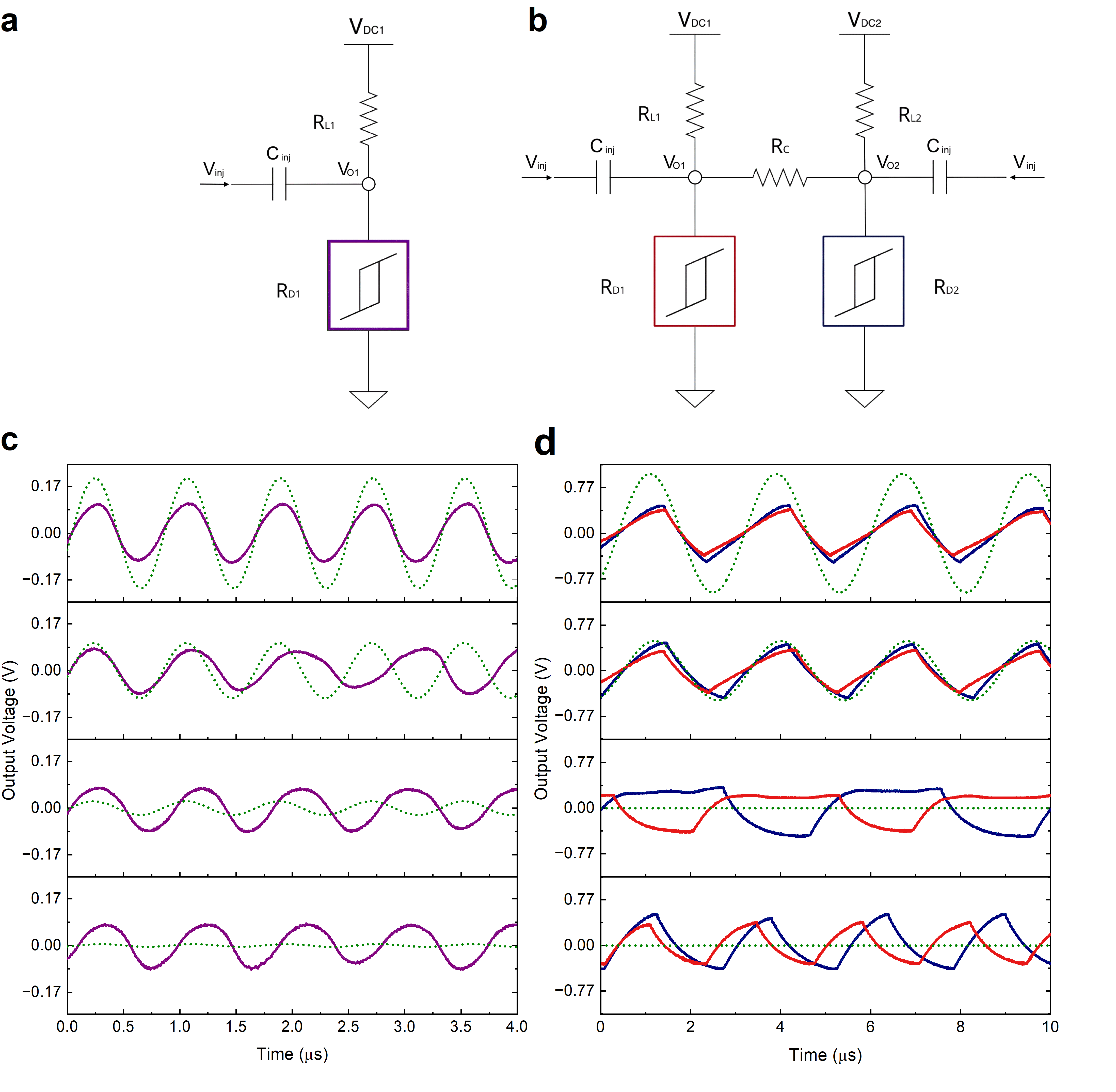}
\caption*{\textbf{Fig. S4 $|$ First Harmonic Injection Locking} When the frequency of the injection locking signal matches the frequency of the oscillator their phases synchronize. Increasing the amplitude of the injection locking signal results in greater phase synchronization. \textbf{a}, The circuit shows the injection locking signal applied to a single CDW oscillator device. \textbf{b}, The circuit shows the injection locking signal applied to two resistively coupled oscillators. \textbf{c}, The first harmonic injection locking (FHIL) (green) of a single oscillator (purple). \textbf{d}, The FHIL of two oscillators (red and blue). (Bottom to top) the oscillators before coupling and before injection locking, after coupling and before injection locking, after coupling and after injection locking with A\textsubscript{inj} $= 1$ V, and A\textsubscript{inj} $= 2$ V.} 
\label{Fig: Figure5.png}
\end{figure}

\newpage
\section*{Supplemental Note 5: Oscillatory Characteristics of a High-Frequency Low Power CDW-QO at RT}
Oscillation frequencies of up to 3.3 MHz are observed in individual 1\emph{T}-TaS\textsubscript{2} devices at room temperature with an amplitude as small as 0.01 V allowing for low power, low latency applications of 1\emph{T}-TaS\textsubscript{2} CDW devices. 
\begin{figure}[h]
\centering
\includegraphics[width=0.8\textwidth]{Figures/SupFigure3.png}
\caption*{\textbf{Fig. S5 $|$} \textbf{Oscillatory Characteristics of a High-Frequency Low Power CDW-QO at RT. a}, The hysteresis in the I-V characteristics of a 1\emph{T}-TaS\textsubscript{2} device at room temperature. A resistive load line from the circuit described in \textbf{2a}, intersects the hysteresis. The width of the hysteresis is $\Delta V \sim$ $0.01$ V. \textbf{b}, Oscillations produced by the circuit shown in \textbf{a}. These oscillations only occur when the voltage across the device intersects the hysteresis as shown in \textbf{a}. \textbf{c}, The frequency of stable oscillations changes as the applied DC bias is adjusted. \textbf{d}, FFT of the highest frequency stable oscillations where $f \sim 3.30$ MHz.} 
\label{Fig: SupFigure3.png}
\end{figure}

\newpage
\section*{Supplemental Note 6: Proof of Theorem. \ref{theorem: Lyapnov funtcion}}

\begin{theorem}{\label{theorem: Lyapnov funtcion}}
Let $\bm{\phi}(t) = 
\begin{bmatrix}
    \phi_1(t) & \phi_1(t) & \cdots & \phi_N(t)
\end{bmatrix}^{\transpose}$, {a choice of} {energy function}  is:
 \begin{align}{\label{eq: energy function}}
        E(\bm{\phi}(t)) = - \frac{1}{2} \sum_{i \neq j } J_{ij}\cos(\phi_{i}(t) -\phi_{j}(t))-\frac{1}{2}K \sum_{i=1}^{N}\sin(2\phi_i(t)+\alpha_1),
        \tag{S1}
    \end{align}
{
    which satisfies
$
        \frac{\partial{E(\bm{\phi}(t))}}{\partial{\bm{\phi}(t)}} = - \dot{\bm{\phi}}(t).
  $
    }
    \end{theorem}
Because $\frac{\partial E(\bm{\phi})}{\partial \bm{\phi}} = \begin{bmatrix} 
    \frac{\partial E(\bm{\phi})}{\partial \phi_1} & \cdots &  \frac{\partial E(\bm{\phi})}{\partial \phi_N}
    \end{bmatrix}$, we differentiate $E(\bm{\phi})$ over each element $\phi_k$:
    {{
    \begin{align*}
        \frac{\partial E(\bm{\phi})}{\partial(\phi_{k})} & =   -\frac{1}{2}\sum_{l=1, l\neq k}^{N} J_{kl} \frac{\partial }{\partial \phi_k(t)}\cos(\phi_{k}(t)-\phi_{l}(t)) - \frac{1}{2}\sum_{l=1,l \neq k }^{N} J_{lk} \frac{\partial }{\partial \phi_k(t)}\cos(\phi_{l}(t)-\phi_{k}(t)) \\
         & - \frac{1}{2} K_s\frac{\partial }{\partial \phi_k(t)}\sin(2\phi_k(t)+\alpha_1) \\
         & = \frac{1}{2}\sum_{l=1}^{N} {J_{kl}} \sin(\phi_{k}(t)-\phi_{l}(t)) - \frac{1}{2}\sum_{l=1}^{N} {J_{lk}} \sin(\phi_{l}(t)-\phi_{k}(t)) - K_s \cos(2 \phi_k(t)+\alpha_1) \\
         & = \sum_{l=1}^{N} \frac{1}{R_{kl}} \sin(\phi_{k}(t)-\phi_{l}(t))  - K_s \cos(2 \phi_k(t)+\alpha_1) \\
         & = - \dot{\phi}_k(t).
    \end{align*}}}
    Since $\dot{{\bm\phi}}(t) = \begin{bmatrix}
        \dot{\phi}_1(t) & \dot{\phi}_2(t) & \cdots & \dot{\phi}_N(t)
    \end{bmatrix}^{\transpose} $, we have
    \begin{align}
        \frac{\partial E(\bm{\phi}(t))}{\partial t} =  \frac{\partial E(\bm{\phi})}{\partial \bm \phi} \dot{\bm{\phi}}(t) = \sum_{k=1}^{N}
        \frac{\partial E(\bm{\phi})}{\partial \phi_{k}} \dot{\phi}_k(t) = - \sum_{k=1}^{N}
        \Big(\dot{\phi}_{k}(t)\Big)^2 \leq 0.
        \tag{S2}
    \end{align}
    Therefore equation \eqref{eq: energy function} is the energy function the coupled CDW oscillators minimize over time under sub-harmonic injection locking. 


\bibliographystyle{plain}